# Study of polyelectrolyte capsule electrophoresis

Anatoly N. Filippov[1],*, Daria Yu. Khanukaeva[1], Petr A. Aleksandrov[2]

[1]*Gubkin University, 65-1 Leninsky prospect, Moscow, 119991 Russia*

[2]*DECO-Geophysical Software Company,*

*1-77 Leninskiye Gory, Moscow, 119992 Russia*

**Abstract.** –This paper investigates the problem of electrophoretic motion of a polyelectrolyte capsule with a porous arbitrary charged conducting shell in an electrolyte (of the same type as the one inside the capsule's cavity) under the action of an external electric field. The corresponding boundary value problem for the velocity components and pressure in the case of small electrical potentials is analytically solved in quadratures. The solution is analyzed numerically for different values of the specific permeability of the capsule, and the thickness of the porous and the electric double layers. The minimum of electrophoretic velocity dependence on the inverse permeability of the porous layer has been found. This means that the capsule can move with the same velocity at different permeabilities of the porous shell, other conditions being equal, which is relevant for the practical use of the discovered effect. It is shown that the electrophoretic mobility decreases upon decrease in the conductivity of the material of constituting the porous layer. This means that a dielectric capsule can be used for electrophoresis as well. Moreover, its velocity will be even greater than that of a conducting capsule, all other conditions being equal.

---

*Corresponding author.

E-mail: filippov.a@gubkin.ru .

# 1. Introduction

The problems of electro-, diffusion-, or thermophoresis, i.e., the motion of particles in a liquid under the influence of an electric field, concentration or temperature gradients respectively, have received a new round of development, despite the fact that the available solutions are already widely used in practice [1-8]. Recent studies often focus on the electrophoresis of colloidal particles of different types [9-20]. The most extensively studied systems are colloidal particles with a grafted layer of polymer molecules. Ohshima [9] considered in detail the rate of electrophoresis of particles with a polyelectrolyte shell. The effect of hydrophobic slipping on the rate of electrophoretic motion of spherical particles were evaluated in [14]. The effect of an induced charge on electrokinetic phenomena was considered in [16-19].

In this paper, we consider the electrophoretic motion of a polyelectrolyte microcapsule with porous walls that are permeable to the electrolyte solution, in the infinite volume of which it is placed. Such objects are used for the synthesis of particles, the production of markers, medicine containers, anticorrosive reagents, etc. [20, 21]. The hydrodynamic behavior of such particles is crucial for the processes of synthesis and practical use of microcapsules since it controls the hydrodynamic flows and determines the specifics of their interaction with the solid boundaries.

We study a microcapsule synthesized from a polyelectrolyte, i.e., a polymer capable of dissociation in solution. As such, it can carry a fixed electric charge, which determines its behavior in an electric field. Traditional models consider solid particle electrophoresis. The electrophoretic behavior of solid particles with a polyelectrolyte shell was studied in [2-4, 19, 22]. If the microcapsule is hollow and filled with an electrolyte solution, its flow significantly affects the movement of the entire capsule [23-25]. In [23], the problem of the flow around a capsule filled with a liquid identical to the dispersion medium is solved; in [24], a capsule containing a liquid that does not mix with the carrier medium is considered; [25] studies the flow around a porous capsule, inside which there is an aggregate with a solid impermeable core and a fractal external structure. In [26], the motion of a polyelectrolyte capsule suspended in an electrolyte under an external electric field is considered. This study took into account that the electrolyte penetrates through a porous capsule, inside which it has the same composition as outside a capsule. The mobility of ions in a liquid was taken to be the same as in a porous medium. One of the most recent papers on this topic is [27], which considers polyelectrolyte capsules, classified into hard and soft ones. A hard capsule refers to a structure with a completely impermeable core, and a soft one has a completely permeable core. In addition, the study considered capsules of an intermediate type – those with a partially permeable core. However, the mathematical formulation of the problem was significantly simplified in [27], namely, the coordinate system was replaced by a one-dimensional Cartesian coordinate system. This simplification was possible since the authors used the linear Debye-Hückel approximation, and the geometry of the problem had spherical symmetry.

In this paper, we use a more general formulation of the problem in spherical coordinates. The motion of the electrolyte is modeled in the framework of the Stokes-Brinkman equations, while taking the action of electrostatic forces into account. An analytical solution is obtained for an arbitrary distribution of the stationary charge density in the capsule material. The electrophoretic mobility of the microcapsule is calculated as a function of the charge density, the conductive properties of the electrolyte and the capsule material, the permeability of the capsule material, and its geometric characteristics.

## 2. Statement of the problem

Let us consider a spherical capsule of radius $b$ with an internal cavity of radius $a$ (Fig. 1). A dry capsule in a neutral external environment (in air) is not charged. We place the capsule into an aqueous solution of the electrolyte, which is originally electroneutral. Then, the dissociation of ionogenic groups occurs on the surface of the capsule pores, and the porous shell of the capsule acquires a volume charge with a density $\rho_f$. The charge that has left the surface of the porous medium contributes to the bulk charge density of the electrolyte $\rho_e$, which will cease to be zero and is determined by the concentration $n_i$ of ions of kind $i$ and their charges $ez_i$, taking into account the sign ($z_i$ – valence of ions of kind $i$, $e$ – electron charge):

$$\rho_e(\mathbf{r}) = \sum_i e\, z_i n_i(\mathbf{r}). \tag{1}$$

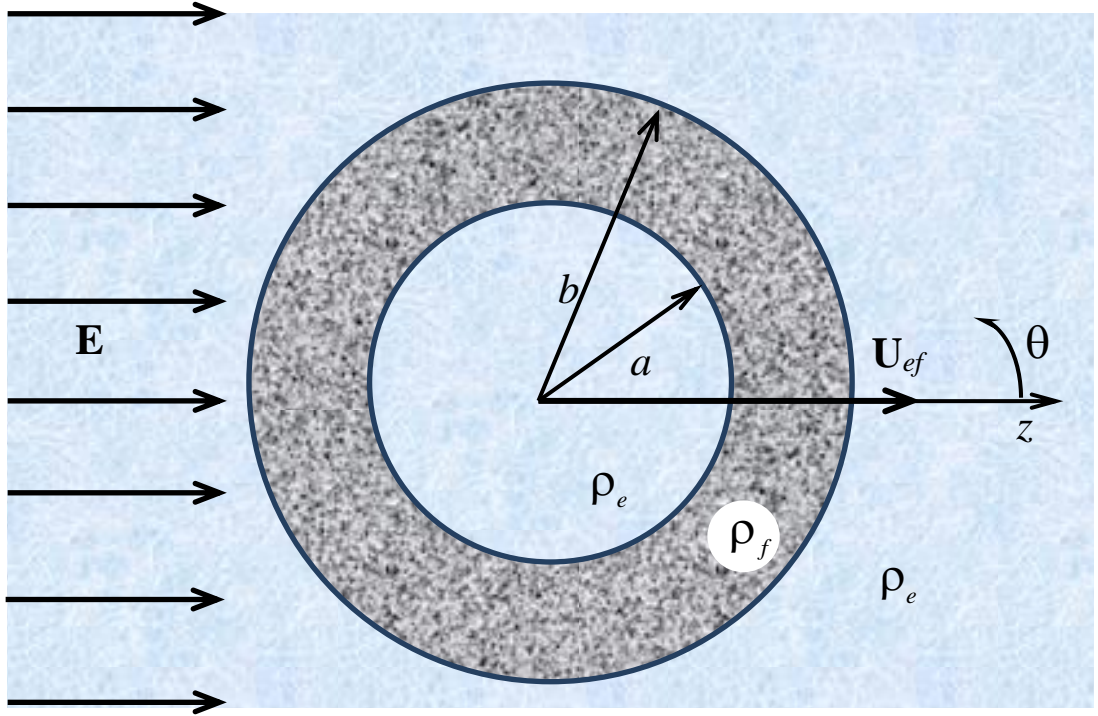

Fig. 1. Polyelectrolyte capsule in an electrolyte solution

On both sides of the inner and outer surfaces of the capsule, electric double layers (EDL) and a certain distribution of the electrostatic potential $\psi_0$ are formed. The equilibrium concentrations $n_i$ of ions in the electrolyte and the distribution of the electrostatic potential $\psi_0$ are obtained from the system of Poisson-Boltzmann equations:

$$n_i(\mathbf{r}) = n_i^\infty \exp\left(-\frac{e\, z_i \psi_0(\mathbf{r})}{k_B T}\right), \tag{2}$$

$$\Delta\psi_0(\mathbf{r}) = -\frac{1}{\varepsilon\varepsilon_0}\begin{cases} \rho_e(\mathbf{r}), & r < a;\ r > b, \\ \rho_e(\mathbf{r}) + \rho_f(\mathbf{r}), & a < r < b, \end{cases} \tag{3}$$

where $k_B$ is the Boltzmann constant, $T$ is the absolute temperature, $n_i^\infty$ is the concentration of ions away from the capsule, $\varepsilon$ is the permittivity of the electrolyte, $\varepsilon_0$ is the electrical constant. When an external electric field is applied to the system, an addition $\psi_1$ to the electrostatic potential $\psi_0$ appears, which is determined by the Laplace equation,

$$\Delta \psi_1 = 0 . \tag{4}$$

The solution of equation (4) is well known. In the case of small potentials $\psi_0$, the system of equations (1)-(3) can be linearized and solved independently of the hydrodynamic part of the problem. The gradient of the total potential $\psi = \psi_0 + \psi_1$ determines the electrostatic force acting on the mobile charges (ions) in the system. We suppose that the capsule is not affected by external forces of different nature. Then, when a uniform electric field is applied to the system, the capsule will start moving, and after a while it will move at a certain constant velocity [1] (electrophoretic velocity).

It should be noted that the electric field will also act on the free ions distributed in the electrolyte. However, we will not explicitly consider the resulting ion transport caused by electrical conductivity, as well as ion diffusion. Nevertheless, the charge transfer they initiate will be is taken into account by introducing the electrical conductivity of the electrolyte along with the electrical conductivity of the capsule, which provide the presence of an electric current in the system when the external electric field is applied.

For the convenience of calculations, we will transfer to a coordinate system that is rigidly connected to the microcapsule. In this system, we have a stationary motion of the electrolyte relative to the capsule. The velocity of the electrolyte away from the microcapsule (at the infinity) is the same (up to the sign) as the electrophoretic velocity $\mathbf{U}_{ef}$.

The equations for the liquid velocity $\mathbf{u}(\mathbf{r})$ in the presence of the electric field having potential $\psi$ are described by the Stokes-Brinkman system:

$$\mu\Delta\mathbf{u} = \nabla p + \rho_e\nabla\psi, \quad 0 < r < a; \quad b < r < \infty,$$

$$\mu\Delta\mathbf{u} - \frac{\mu}{k}\mathbf{u} = \nabla p + \rho_e\nabla\psi, \quad a < r < b, \tag{5}$$

$$\text{div}\,\mathbf{u} = 0,$$

where $p$ is the pressure, $\mu$ is the viscosity of the electrolyte, which will be considered the same in all regions of the flow, $k$ is the permeability of the porous medium. Note that in all regions of the flow, the electrostatic force equals the product of the gradient of the electric field potential and the charge density of the electrolyte, i.e., only mobile ions are taken. When composing the equation of fluid motion, we are not interested in the force acting on the static charges of the density $\rho_f$.

The boundary conditions in the spherical coordinate system $(r, \theta, \phi)$ associated with the capsule have the form,

$$|\mathbf{u}| \leq M = \text{const} \text{ at } r \to 0,$$

$$\mathbf{u} \to -\mathbf{U}_{\text{ef}} \quad \text{at } r \to \infty, \tag{6}$$

where $\mathbf{U}_{\text{ef}}$ – electrophoretic velocity, which is the required quantity in this problem.

We also have continuity conditions for the velocity vector components on the inner and outer surfaces of the capsule:

$$u_r\big|_{r=a_-} = u_r\big|_{r=a_+},$$

$$u_r\big|_{r=b_-} = u_r\big|_{r=b_+},$$

$$u_\theta\big|_{r=a_-} = u_\theta\big|_{r=a_+}, \tag{7}$$

$$u_\theta\big|_{r=b_-} = u_\theta\big|_{r=b_+}.$$

Similar continuity conditions should be written for the components of the total stress tensor equal to the sum of the hydrodynamic stress tensor,

$$\mathbf{\Pi} = -p\mathbf{G} + \mu\left(\nabla\mathbf{u} + (\nabla\mathbf{u})^T\right) \tag{8}$$

and the Maxwell tensor,

$$\mathbf{\Pi}^M = \varepsilon\varepsilon_0\left(\nabla\psi\nabla\psi - \frac{1}{2}(\nabla\psi \cdot \nabla\psi)\mathbf{G}\right), \tag{9}$$

where superscript $T$ denotes transposition, $\mathbf{G}$ is the metric tensor. We assume for simplicity that the dielectric permittivity of the electrolyte equals to that of the capsule impregnated with the electrolyte. In addition, we accept that the electrostatic potential is continuous together with its derivatives. This makes it possible to ignore the Maxwell stress tensor in the boundary conditions, since it is automatically continuous in this case. As a result, for the hydrodynamic stress tensor respectively, we have

$$
\begin{aligned}
\Pi_{rr}\big|_{r=a_-} &= \Pi_{rr}\big|_{r=a_+}, \\
\Pi_{rr}\big|_{r=b_-} &= \Pi_{rr}\big|_{r=b_+}, \\
\Pi_{r\theta}\big|_{r=a_-} &= \Pi_{r\theta}\big|_{r=a_+}, \\
\Pi_{r\theta}\big|_{r=b_-} &= \Pi_{r\theta}\big|_{r=b_+}.
\end{aligned}
\tag{10}
$$

It should be mentioned that the difference between the dielectric permittivity of the capsule and the fluid medium should not make significant changes in the obtained results. This is since the force acting on the dielectric particles in a uniform external electrostatic field is zero. That is, with the emerging dipole nature of the potential distribution in the vicinity of the capsule, the total contribution of Maxwell stresses gives zero force. In the last case, the electrostatic potential is similar to the field created by the dipole, which suggests a zero contribution from the Maxwell stress tensor to the total force acting on the particle.

The velocity of electrophoresis is found from the condition that the sum of the electrostatic $\mathbf{F}_e$ and hydrodynamic $\mathbf{F}_h$ forces acting on the capsule is equal to zero:

$$
\begin{aligned}
\mathbf{F}_e &= -\int_V \rho_f \nabla(\psi_0 + \psi_1)\,dV, \\
\mathbf{F}_h &= \oint_S \mathbf{\Pi} \cdot \mathbf{n}\,dS,
\end{aligned}
\tag{11}
$$

where the integration is extended over the volume $V$ of the porous layer and over its outer surface $S$, respectively.

## 3. Distribution of the electrostatic potential

## 3.1. Capsule in a bulk electrolyte solution

We will find a solution to the problem (1)-(3) in a linear approximation, limiting ourselves to the case of small potentials. Due to the spherical symmetry of the problem, $\psi_0$ is only a function of the radial coordinate. Taking this fact into account, equations (1)-(3) can be reduced to the following equations:

$$\frac{1}{r^2}\frac{d}{dr}\left(r^2\frac{d\psi_0}{dr}\right) = -\frac{1}{\varepsilon\varepsilon_0}\begin{cases}\sum_i e\,z_i n_i^\infty \exp\left(-\frac{e\,z_i\psi_0}{k_BT}\right)+\rho_f(r), & a<r<b,\\[2mm] \sum_i e\,z_i n_i^\infty \exp\left(-\frac{e\,z_i\psi_0}{k_BT}\right), & r<a;\quad b<r.\end{cases} \tag{12}$$

The condition of the electrolyte electroneutrality away from the capsule reads $\sum_i e\,z_i n_i^\infty = 0$. Therefore, after linearization, equation (12) will take the form,

$$\frac{1}{r^2}\frac{d}{dr}\left(r^2\frac{d\psi_0}{dr}\right) = \begin{cases}\kappa^2\psi_0 - \dfrac{\rho_f(r)}{\varepsilon\varepsilon_0}, & a<r<b,\\[3mm] \kappa^2\psi_0, & r<a;\quad b<r,\end{cases} \tag{13}$$

where the reverse Debye radius is introduced,

$$\kappa = \left(\frac{1}{\varepsilon\varepsilon_0 k_BT}\sum_{i=1}^N z_i^2 e^2 n_i^\infty\right)^{1/2}. \tag{14}$$

Let us turn on to the following dimensionless variables and quantities:

$$\tilde{r}=\frac{r}{b},\quad \frac{a}{b}=\gamma,\quad \tilde{\kappa}=\kappa\cdot b,\quad \tilde{\nabla}=\nabla\cdot b,\quad \tilde{\Delta}=\Delta\cdot b^2,\quad \tilde{\psi}_0=\frac{\psi_0\cdot\kappa}{E},\quad \tilde{\rho}_{f0}=\frac{\rho_f}{\rho_0},$$

where $E$ is the strength of the applied external electric field, $\rho_0$ is the characteristic value of the volume density of charges induced on the capsule, which is 1 mmol/L = 1 mol/m³. Thus, it makes sense to vary the value of the dimensionless charge density in the range (0; 10). Note that for the introduced geometry, we have the value $\gamma\in(0;1)$, and for the contemporary capsules it is at least 0.9. The characteristic values of Debye's radii for electrolytes are of the order of $1\div10$ nm, polyelectrolyte capsules have characteristic sizes of the order of $1 \div 5\,\mu m$. Therefore, $\tilde{\kappa} \sim 10^2 \div 5 \cdot 10^3$. Note that for the linear approximation to be applicable, the following value must remain small: $\dfrac{e\, z_i \psi_0}{k_B T} \sim \dfrac{1,6 \cdot 10^{-19} \psi_0}{1,4 \cdot 10^{-23} \cdot 300} \sim \dfrac{\psi_0}{0,03}$, i.e., $\psi_0 \sim 3\,mV$. The characteristic values of the external field strengths used in electrophoresis are of the order of $10^2 \div 10^3$ V/m. Therefore, the introduced dimensionless quantity has the order

$$\tilde{\psi}_0 \sim \frac{3 \cdot 10^{-3} V}{\left(10^2 \div 10^3 V / m\right) \cdot \left(10^{-9} \div 10^{-8} m\right)} \sim 300 \div 30000.$$

Equation (13) in dimensionless variables takes the form,

$$\frac{1}{r^2} \frac{d}{dr}\left( r^2 \frac{d\psi_0}{dr} \right) = \begin{cases} \kappa^2\left(\psi_0 - \rho(r)\right), & r \in (1;\gamma), \\ \kappa^2 \psi_0, & r \notin (1;\gamma), \end{cases} \tag{15}$$

where the tildes over all dimensionless variables and quantities are omitted for simplicity and the following dimensionless complex is introduced: $\rho = \tilde{\rho}_{f0}\,\dfrac{\rho_0}{\varepsilon \varepsilon_0 E \kappa}$.

The solutions of equations (15) under the conditions of continuity of the electric potential and its derivatives have the form:

at $r < \gamma$,

$$\psi_{0a}(r) = \frac{\rho}{\kappa}\Big[ (1+\kappa\gamma)\exp\left(-\kappa\gamma\right) - (1+\kappa)\exp\left(-\kappa\right) \Big]\frac{\mathrm{sh}(\kappa r)}{r}; \tag{16}$$

at $\gamma < r < 1$,

$$\psi_{0ab}(r) = \frac{\rho}{\kappa}\Bigg\{ \Big[ (1+\kappa\gamma)\exp\left(-\kappa\gamma\right) - (1+\kappa)\exp\left(-\kappa\right) \Big]\frac{\mathrm{sh}(\kappa r)}{r} + \kappa\left[ 1 - \frac{\gamma}{r}\mathrm{ch}(\kappa(r-\gamma)) - \frac{1}{\kappa r}\mathrm{sh}(\kappa(r-\gamma)) \right] \Bigg\}; \tag{17}$$

at $1 < r$,

$$\psi_{0b}(r) = \frac{\rho}{\kappa}\Big[ \kappa\,\mathrm{ch}(\kappa) - \kappa\gamma\,\mathrm{ch}(\kappa\gamma) - \mathrm{sh}(\kappa) + \mathrm{sh}(\kappa\gamma) \Big]\frac{\exp\left(-\kappa r\right)}{r}. \tag{18}$$

### 3.2. Capsule in an external electric field

We assume that when an external electric field of the medium intensity is applied, the EDL near the capsule does not polarize, since its polarization in moderate fields contributes a small effect to the electrophoresis rate [1]. We assume that a homogeneous external electric field with a strength $\mathbf{E}$ is directed along the $z$-axis of the introduced coordinate system. The resulting increment $\psi_1$ to the electrostatic potential $\psi_0$ is described by equation (4), which remains unchanged when turning to dimensionless variables $\tilde{r} = \dfrac{r}{b}, \quad \tilde{\psi}_1 = \dfrac{\psi_1 \cdot \kappa}{E}$. And the solution of (4), taking the boundary conditions at $r \to 0$ and $r \to \infty$ into account, has the form:

$$\psi_{1a} = -Cr\cos\theta, \qquad r < \gamma, \tag{19}$$

$$\psi_{1ab} = -(Br + D\frac{1}{r^2})\cos\theta, \qquad \gamma < r < 1, \tag{20}$$

$$\psi_{1b} = -(r + A\frac{1}{r^2})\cos\theta, \qquad 1 < r, \tag{21}$$

where the angle $\theta$ is measured from the positive direction of the $z$-axis.

Constants $A$, $B$, $C$, $D$ can be determined from the conditions of continuity of the potential $\psi_{1a}\big|_{r=\gamma} = \psi_{1ab}\big|_{r=\gamma}, \quad \psi_{1ab}\big|_{r=1} = \psi_{1b}\big|_{r=1}$ and continuity of the electric current density $\mathbf{I} = \sigma^* \cdot \mathbf{E} = -\sigma^* \cdot \nabla\psi_1$ in the normal direction to the inner and outer surfaces of the capsule: $\sigma_e \dfrac{\partial \psi_{1a}}{\partial r}\bigg|_{r=\gamma} = \sigma_i \dfrac{\partial \psi_{1ab}}{\partial r}\bigg|_{r=\gamma}, \quad \sigma_i \dfrac{\partial \psi_{1ab}}{\partial r}\bigg|_{r=1} = \sigma_e \dfrac{\partial \psi_{1b}}{\partial r}\bigg|_{r=1},$ where $\sigma^*$ is the medium conductivity. At the same time, we believe that the conductivity $\sigma_e$ of the electrolyte and that of the capsule filled with the electrolyte $\sigma_i$ are different. We will operate with the dimensionless ratio of these conductivities $\sigma = \sigma_i / \sigma_e$, which can be less or more than unity. After formal calculations, we find,

$$A = \frac{(1-\sigma)(1+2\sigma)(1-\gamma^3)}{2 - 2\gamma^3(1-\sigma)^2 + 5\sigma + 2\sigma^2}, \tag{22}$$

$$B = \frac{3+6\sigma}{2 - 2\gamma^3(1-\sigma)^2 + 5\sigma + 2\sigma^2}, \tag{23}$$

$$C = \frac{9\sigma}{2 - 2\gamma^3(1-\sigma)^2 + 5\sigma + 2\sigma^2}, \tag{24}$$

$$D = \frac{3(\sigma-1)\gamma^3}{2 - 2\gamma^3(1-\sigma)^2 + 5\sigma + 2\sigma^2}. \tag{25}$$

As can be seen from the solution (19)-(21), as well as from Fig. 2a,d, the dependence $\psi_1(r)$ does not differ much from the linear one. The value of $\sigma$ mainly affects the slope and curvature of the dependence $\psi_1(r)$ at the intersection of the boundaries of the porous layer $r = \gamma, \quad r = 1$. The type of dependence $\psi_0(r)$ is determined by the amount of charge induced on the capsule, the thickness of the EDL, characterized by the value of the Debye radius $1/\kappa$, and the thickness of the porous wall of the capsule. Obviously, the distribution $\psi_0(r)$ has a maximum in the region of charge concentration $\gamma < r < 1$, which value is proportional to the ratio of the charge density to the external field strength and is characterized by the value $\rho$, and the blur measure of the maximum depends on $\kappa$. For sufficiently large values of $\kappa$ (thin EDL), the potential distribution $\psi_0(r)$ is almost stepwise, as can be seen in Fig. 2b. When choosing physically correct values of the problem parameters, the value of $\psi_1(r)$ represents a small correction to $\psi_0(r)$, without fundamentally changing the nature of $\psi_0(r)$ dependence. Characteristic curves $\psi_0(r)$, $\psi_1(r)$, and $\psi_0(r) + \psi_1(r)$ are shown in Fig. 2 for two sets of the parameter values that differ in the conductivity ratio $\sigma$ and the thickness of the porous layer (the location $\gamma$ of the inner wall of the capsule). When drawing the curves of Fig. 2a,b,c, we used $\gamma = 0.5;\quad \sigma = 10$, while for Fig. 2d,e,f we used $\gamma = 0.99;\quad \sigma = 1$, the other parameters had the following values: $\rho = 10^3;\quad \kappa = 30;\quad \theta = 0$.

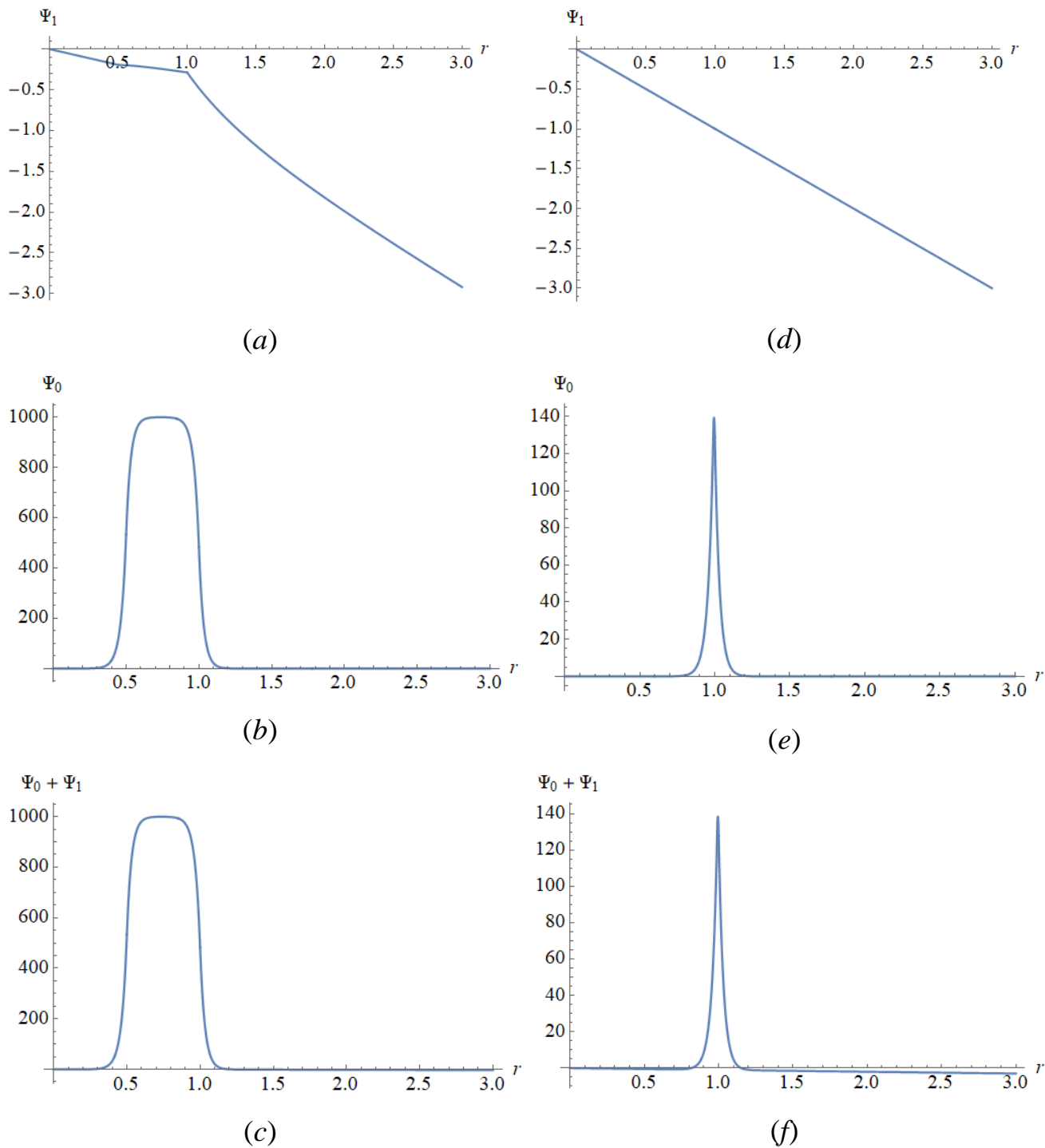

(a)

(d)

(b)

(e)

(c)

(f)

Fig. 2. Dependences of electric potentials $\psi_0(r)$, $\psi_1(r)$ and $\psi_0(r) + \psi_1(r)$: (*a, b, c*) – $\rho$=1000; $\gamma$=0.5; $\kappa$=30; $\sigma$=10; $\theta$=0; (*d, e, f*) – $\rho$=1000; $\gamma$=0.99; $\kappa$=30; $\sigma$=1; $\theta$=0.

## 4. Velocity and pressure fields

Before proceeding to the solution of the hydrodynamic problem, let us convert the system of equations (5) to a dimensionless form. We refer the flow velocity to a characteristic velocity $U_0$, and the pressure to a characteristic pressure $P_0$, which we define so that $\dfrac{E^2 \varepsilon \varepsilon_0}{\kappa \mu U_0} = 1$ and $\dfrac{P_0 b}{\mu U_0} = 1$, i.e., $U_0 = \dfrac{E^2 \varepsilon \varepsilon_0}{\kappa \mu}$, $P_0 = \dfrac{\mu U_0}{b} = \dfrac{E^2 \varepsilon \varepsilon_0}{\kappa b}$.

Then the dimensionless Stokes equation will take the form

$$\Delta \mathbf{u} = \nabla p + \frac{\rho_\varepsilon}{\rho_1} \kappa \nabla \psi, \quad r \notin (\gamma; 1), \qquad (26)$$

where $\rho_\varepsilon = \dfrac{\rho_e}{\rho_0}$ is the dimensionless charge density of the electrolyte solution, and the dimensionless complex $\rho_1 = \dfrac{\varepsilon \varepsilon_0 E \kappa}{\rho_0}$ is introduced. Taking the characteristic values of the external electric field strength, the reverse Debye radius, and the charge density of the capsule into account, we can evaluate an order of magnitude $\rho_1 = \dfrac{80 \cdot 9 \cdot 10^{-12} \cdot 10^2 \div 10^3 \cdot 10^8 \div 10^9}{10^5} \sim 10^{-4} \div 10^{-2}$. It should be noticed that while deriving equation (13), it is obtained that $\dfrac{\rho_e}{\varepsilon \varepsilon_0} = -\kappa^2 \psi_0$ (all values are dimensional). Therefore, in a dimensionless form we have $\rho_\varepsilon = -\rho_1 \psi_0$ (all quantities are dimensionless). Thus, having the solution (16)-(18), we automatically obtain the distribution of the charge density $\rho_\varepsilon(r)$ in the electrolyte. It, together with the known solutions for $\psi_0(r)$ and $\psi_1(r, \theta)$, forms the "right-hand side" of equation (26). The mentioned term will also appear in the dimensionless Brinkman equation, in which the parameter $s^2 = b^2 / k$ will still emerge:

$$\Delta \mathbf{u} - s^2 \mathbf{u} = \nabla p + \frac{\rho_\varepsilon}{\rho_1} \kappa \nabla \psi, \quad r \in (\gamma; 1). \qquad (27)$$

The dimensionless form of the continuity equation coincides with the dimensional one:

$$\nabla \cdot \mathbf{u} = 0. \tag{28}$$

### 4.1. Pressure field

Since the right-hand sides of equations (26) and (27) consist of two terms, one of which ($\psi_0$) depends only on $r$, and the other ($\psi_1$) depends on $r$ and $\theta$, the partial solution of these inhomogeneous equations will also be represented as the sum of the two terms. One of them will be associated with $\psi_0$ and depend only on $r$, and the other, associated with $\psi_1$, will depend on $r$ and $\theta$. In addition, it will be necessary to add a known solution of the corresponding homogeneous equations [23] to these expressions.

Thus, we have $p = p_0(r) + p_1(r, \theta)$, $\mathbf{u} = \mathbf{u}_0(r) + \mathbf{u}_1(r, \theta)$, with $\mathbf{u}_0 = \mathbf{0}$, since the field of stationary distribution of charges in the electrolyte does not cause the displacement of the capsule, but only leads to a certain distribution of pressure. The equations of motion for quantities that depend only on $r$ will take the following form in all flow regions:

$$0 = \nabla p_0 + \frac{\rho_\varepsilon}{\rho_1} \kappa \nabla \psi_0.$$

Since $\rho_\varepsilon = -\rho_1 \psi_0$, and the solution for $\psi_0$ is known and is given by formulae (16)-(18), we obtain,

$$p_0(r) = \begin{cases} -\int_0^r \kappa \dfrac{\rho_\varepsilon(r')}{\rho_1} \dfrac{\partial \psi_{0a}}{\partial r'} dr' = \kappa \int_0^r \psi_{0a} \dfrac{\partial \psi_{0a}}{\partial r'} dr', & r < \gamma \\[2mm] -\int_\gamma^r \kappa \dfrac{\rho_\varepsilon(r')}{\rho_1} \dfrac{\partial \psi_{0ab}}{\partial r'} dr' = \kappa \int_\gamma^r \psi_{0ab} \dfrac{\partial \psi_{0ab}}{\partial r'} dr', & \gamma < r < 1 \\[2mm] \int_r^\infty \kappa \dfrac{\rho_\varepsilon(r')}{\rho_1} \dfrac{\partial \psi_{0b}}{\partial r'} dr' = -\kappa \int_r^\infty \psi_{0b} \dfrac{\partial \psi_{0b}}{\partial r'} dr', & r > 1 \end{cases} \tag{29}$$

For unknowns with index 1, we have

$$\Delta \mathbf{u}_1 = \nabla p_1 + \frac{\rho_\varepsilon}{\rho_1} \kappa \nabla \psi_1, \quad r \notin (\gamma; 1),$$

$$\Delta \mathbf{u}_1 - s^2 \mathbf{u}_1 = \nabla p_1 + \frac{\rho_\varepsilon}{\rho_1} \kappa \nabla \psi_1, \quad r \in (\gamma; 1).$$

Taking the divergence of these equations and considering the identity $\mathrm{div}(\Delta \mathbf{u}) = \Delta(\mathrm{div}\,\mathbf{u})$ together with equation (28), we obtain an equation for determining the pressure field $p_1$, which is valid in all three flow regions:

$$\Delta p_1 + \mathrm{div}\left(\frac{\rho_\varepsilon}{\rho_1} \kappa \nabla \psi_1\right) = 0.$$

We can transform the second term in the previous equation as follows:

$$\kappa \, \mathrm{div}\left(\frac{\rho_\varepsilon}{\rho_1} \nabla \psi_1\right) = \kappa \left(\nabla\left(-\psi_0\right) \cdot \nabla \psi_1 + \frac{\rho_\varepsilon}{\rho_1} \underset{=0}{\Delta \psi_1}\right) =$$

$$= -\kappa \begin{pmatrix} \partial \psi_0 / \partial r \\ 0 \\ 0 \end{pmatrix} \cdot \begin{pmatrix} \partial \psi_1 / \partial r \\ (\partial \psi_1 / \partial \theta) / r \\ 0 \end{pmatrix} = -\kappa \frac{\partial \psi_0}{\partial r} \frac{\partial \psi_1}{\partial r}.$$

Calculating the derivatives of the known functions $\psi_0$ and $\psi_1$, we obtain the equations for finding $p_1$ in each of the flow regions:

$$\Delta p_1 = \begin{cases} -\kappa C \cos\theta \dfrac{\partial \psi_{0a}}{\partial r}, & r < \gamma, \\[2mm] -\kappa (B - 2D \dfrac{1}{r^3}) \cos\theta \dfrac{\partial \psi_{0ab}}{\partial r}, & \gamma < r < 1, \\[2mm] -\kappa (1 - 2A \dfrac{1}{r^3}) \cos\theta \dfrac{\partial \psi_{0b}}{\partial r}, & 1 < r. \end{cases} \qquad (30)$$

The right-hand sides of equations (30) are known: the constants $A$, $B$, $C$, $D$ are determined by the relations (22)-(25), respectively, and the expressions for $\psi_0$ are determined by the relations (16)-(18). The solution of equations (30) will be found in the form,

$$p_1(r, \theta) = -Q(r) \cos\theta + Const,$$

then for $Q$ we get the following equations,

$$\frac{\partial}{\partial r}\left(\frac{1}{r^2}\frac{\partial}{\partial r}(r^2 Q)\right) = \begin{cases} \kappa C \dfrac{\partial \psi_{0a}}{\partial r}, & r < \gamma, \\[2ex] \kappa (B - 2D \dfrac{1}{r^3})\dfrac{\partial \psi_{0ab}}{\partial r}, & \gamma < r < 1, \\[2ex] \kappa (1 - 2A \dfrac{1}{r^3})\dfrac{\partial \psi_{0b}}{\partial r}, & 1 < r. \end{cases} \quad (31)$$

We need to mention that $\psi_0$ satisfies the Laplace equation, i.e. $r^2 \kappa^2 \psi_0 = \dfrac{\partial}{\partial r}\left(r^2 \dfrac{\partial \psi_0}{\partial r}\right)$. This allows us to integrate equations (31) formally as follows.

In the region $r < \gamma$ we have,

$$\int_0^r \frac{\partial}{\partial r'}\left(\frac{1}{r'^2}\frac{\partial}{\partial r'}(r'^2 Q)\right)dr' = \kappa C \int_0^r \frac{\partial \psi_{0a}}{\partial r'}dr',$$

$$\frac{1}{r^2}\frac{\partial}{\partial r}(r^2 Q) - const_0 = \kappa C[\psi_{0a}(r) - \psi_{0a}(0)],$$

$$\frac{\partial}{\partial r}(r^2 Q) = C\kappa r^2 \psi_{0a}(r) + const_1 r^2 = \frac{C}{\kappa}\frac{\partial}{\partial r}\left(r^2 \frac{\partial \psi_{0a}}{\partial r}\right) + const_1 r^2.$$

Integration between 0 and $r$ gives,

$$r^2 Q - 0 = \frac{C}{\kappa}[r^2 \frac{\partial \psi_{0a}}{\partial r} - 0] + const_1 \frac{r^3}{3} - 0.$$

As a result, we get,

$$Q = \frac{C}{\kappa}\frac{\partial \psi_{0a}}{\partial r} - K_1 r, \quad (32)$$

where the integration constant is re-designated as $K_1$.

Carrying out similar calculations in the interval $\gamma < r < 1$, we get,

$$\int_\gamma^r \frac{\partial}{\partial r'}\left(\frac{1}{r'^2}\frac{\partial}{\partial r'}(r'^2 Q)\right)dr' = \kappa B \int_\gamma^r \frac{\partial \psi_{0ab}}{\partial r'}dr' - \kappa 2D \int_\gamma^r \frac{1}{r'^3}\frac{\partial \psi_{0ab}}{\partial r'}dr',$$

$$\frac{1}{r^2}\frac{\partial}{\partial r}(r^2 Q) - const_\gamma =$$

$$= \kappa B[\psi_{0ab}(r) - \psi_{0ab}(\gamma)] - 2\kappa D\left[\frac{\psi_{0ab}(r)}{r^3} - \frac{\psi_{0ab}(\gamma)}{\gamma^3} + 3\int_\gamma^r \frac{\psi_{0ab}(r')}{r'^4}dr'\right],$$

$$\frac{\partial}{\partial r}(r^2 Q) = B\kappa r^2 \psi_{0ab}(r) + const_2 r^2 - 2\kappa D\left[\frac{\psi_{0ab}(r)}{r} + 3r^2\int_\gamma^r \frac{\psi_{0ab}(r')}{r'^4}dr'\right] =$$

$$= \frac{B}{\kappa}\frac{\partial}{\partial r}\left(r^2\frac{\partial\psi_{0ab}}{\partial r}\right) + const_2 r^2 - 2\kappa D\frac{\partial}{\partial r}\left[r^3\int_\gamma^r \frac{\psi_{0ab}(r')}{r'^4}dr'\right].$$

Integration in the range from $\gamma$ to $r$ gives,

$$r^2 Q - \gamma^2 Q(\gamma) = \frac{B}{\kappa}\left(r^2\frac{\partial\psi_{0ab}}{\partial r} - \gamma^2\frac{\partial\psi_{0ab}(\gamma)}{\partial r}\right) + const_2\left(\frac{r^3}{3} - \frac{\gamma^3}{3}\right) -$$

$$- 2\kappa D\left[r^3\int_\gamma^r \frac{\psi_{0ab}(r')}{r'^4}dr'\right] - 0$$

So, in the region $\gamma < r < 1$ we get,

$$Q = \frac{B}{\kappa}\frac{\partial\psi_{0ab}}{\partial r} - 2\kappa D \ r\int_\gamma^r \frac{\psi_{0ab}(r')}{r'^4}dr' - K_2 r - \frac{K_3}{r^2}, \tag{33}$$

where two new constants $K_2$ and $K_3$ are introduced.

Similarly, in the region $r > 1$ we have,

$$\int_r^\infty \frac{\partial}{\partial r'}\left(\frac{1}{r'^2}\frac{\partial}{\partial r'}(r'^2 Q)\right)dr' = \kappa\int_r^\infty \frac{\partial\psi_{0b}}{\partial r'}dr' - 2A\kappa\int_r^\infty \frac{1}{r'^3}\frac{\partial\psi_{0b}}{\partial r'}dr',$$

$$0 - \frac{1}{r^2}\frac{\partial}{\partial r}(r^2 Q) = \kappa[0 - \psi_{0b}(r)] - 2A\kappa\left[\frac{1}{r'^3}\psi_{0b}\Big|_r^\infty + 3\int_r^\infty \frac{1}{r'^4}\frac{\partial\psi_{0b}}{\partial r'}dr'\right],$$

$$-\frac{\partial}{\partial r}(r^2 Q) = -\kappa r^2\psi_{0b}(r) - 2A\kappa\left[3r^2\int_r^\infty \frac{1}{r'^4}\frac{\partial\psi_{0b}}{\partial r'}dr' - \frac{1}{r}\psi_{0b}(r)\right] =$$

$$= -\frac{1}{\kappa}\frac{\partial}{\partial r}\left(r^2\frac{\partial\psi_{0b}}{\partial r}\right) - 2A\kappa\frac{\partial}{\partial r}\left[r^3\int_r^\infty \frac{1}{r'^4}\psi_{0b}dr'\right].$$

After integrating in the range from $r$ to infinity, we get,

$$r^2 Q - const_\infty = \frac{1}{\kappa} r^2 \frac{\partial \psi_{0b}}{\partial r} - 0 + 2A\kappa r^3 \int_r^\infty \frac{\psi_{0b}(r')}{r'^4} dr' - 0.$$

Finally, for the region $r > 1$, we obtain,

$$Q = \frac{1}{\kappa} \frac{\partial \psi_{0b}}{\partial r} + 2A\kappa r \int_r^\infty \frac{\psi_{0b}(r')}{r'^4} dr' - \frac{K_4}{r^2}, \tag{34}$$

where the constant $K_4$ is introduced. Combining solutions (29), (32)-(34), we get the final expressions for the pressure,

$$p(r,\theta) = \begin{cases} \left( K_1 r - \dfrac{C}{\kappa} \dfrac{\partial \psi_{0a}}{\partial r} \right) \cos\theta + \kappa \int_0^r \psi_{0a} \dfrac{\partial \psi_{0a}}{\partial r'} dr' + p_{0a}, & r < \gamma \\[4mm] \left( K_2 r + \dfrac{K_3}{r^2} - \dfrac{B}{\kappa} \dfrac{\partial \psi_{0ab}}{\partial r} + 2\kappa D \, r \int_\gamma^r \dfrac{\psi_{0ab}(r')}{r'^4} dr' \right) \cos\theta + \\[4mm] \qquad\qquad + \kappa \int_\gamma^r \psi_{0ab} \dfrac{\partial \psi_{0ab}}{\partial r'} dr' + p_{0ab}, & \gamma < r < 1 \\[4mm] \left( \dfrac{K_4}{r^2} - \dfrac{1}{\kappa} \dfrac{\partial \psi_{0b}}{\partial r} - 2A\kappa r \int_r^\infty \dfrac{\psi_{0b}(r')}{r'^4} dr' \right) \cos\theta - \\[4mm] \qquad\qquad - \kappa \int_r^\infty \psi_{0b} \dfrac{\partial \psi_{0b}}{\partial r'} dr' + p_{0b}, & r > 1 \end{cases} \tag{35}$$

The constants $K_1 \div K_4$ will be found after determining the velocity components, which we are now going to find.

### 4.2. Velocity field

To determine the velocity, we apply the curl operation to equations (26) and (27), then the term with pressure disappears, and the term containing the electrostatic potential is transformed as follows (we do not account for the multiplier $\kappa / \rho_1$):

$$\nabla \times (\rho_\varepsilon \nabla \psi) = \nabla \rho_\varepsilon \times \nabla \psi + \rho_\varepsilon \underbrace{\nabla \times \nabla \psi}_{=\mathbf{0}} = \underbrace{\nabla \rho_\varepsilon \times \nabla \psi_0}_{=\mathbf{0}, \text{ since } \nabla \rho_\varepsilon \| \nabla \psi_0} + \nabla \rho_\varepsilon \times \nabla \psi_1.$$

The first term in the previous equality has turned to zero, since $\psi_0$ and $\rho_\varepsilon$ depend only on $r$, and their gradients are parallel. For the second term we have,

$$\frac{\kappa}{\rho_1} \nabla \rho_\varepsilon \times \nabla \psi_1 = \frac{\kappa}{\rho_1} \begin{pmatrix} \partial \rho_\varepsilon / \partial r \\ 0 \\ 0 \end{pmatrix} \times \begin{pmatrix} \partial \psi_1 / \partial r \\ (\partial \psi_1 / \partial \theta) / r \\ 0 \end{pmatrix} =$$

$$= -\mathbf{i}_\phi \sin \theta \kappa \begin{cases} C \dfrac{\partial \psi_{0a}}{\partial r}, & r < \gamma, \\ \left( B + \dfrac{D}{r^3} \right) \dfrac{\partial \psi_{0ab}}{\partial r}, & \gamma < r < 1, \\ \left( 1 + \dfrac{A}{r^3} \right) \dfrac{\partial \psi_{0b}}{\partial r}, & 1 < r, \end{cases}$$

where $\mathbf{i}_\phi$ is the unit vector of the spherical coordinate system ($r$, $\theta$, $\phi$).

Based on the symmetry of the problem, we will seek for a solution for the velocity $\mathbf{u}$ in the form:

$$\begin{aligned} u_r &= -\frac{2}{r} h(r) \cos \theta, \\ u_\theta &= \frac{1}{r} \frac{d(rh)}{dr} \sin \theta, \\ u_\phi &= 0, \end{aligned} \tag{36}$$

where $h(r)$ is the required function that depends only on the radius $r$. Then the continuity equation is satisfied identically. The left-hand sides of equations (26) and (27) will also include $\mathrm{curl}\,\mathbf{u}$ and $\mathrm{curl}\,\Delta\mathbf{u}$, which will take the form,

$$\mathrm{curl}\,\mathbf{u} = \mathbf{i}_\phi L(h) \sin \theta,$$

$$\mathrm{curl}\,\Delta\mathbf{u} = \Delta \mathrm{curl}\,\mathbf{u} = \mathbf{i}_\phi L(L(h)) \sin \theta,$$

where the operator $L(.)$ is defined as,

$$L = \frac{d}{dr} \left( \frac{1}{r^2} \frac{d}{dr} r^2 \right). \tag{37}$$

As a result, the equations that need to be solved to find the components of the flow velocity will take the form,

$$LL(h) = -\kappa C \frac{\partial \psi_{0a}}{\partial r}, \qquad\qquad r < \gamma,$$

$$LL(h) - s^2 L(h) = -\kappa \left( B + \frac{D}{r^3} \right) \frac{\partial \psi_{0ab}}{\partial r}, \qquad \gamma < r < 1, \quad (38)$$

$$LL(h) = -\kappa \left( 1 + \frac{A}{r^3} \right) \frac{\partial \psi_{0b}}{\partial r}, \qquad\qquad 1 < r.$$

We find a partial solution of the inhomogeneous equations (38), the right-hand sides of which we denote respectively $G_a(r), G_{ab}(r), G_b(r)$. In the region $r < \gamma$ we have,

$$\frac{d}{dr} \left( \frac{1}{r^2} \frac{d}{dr}(r^2 L h) \right) = G_a(r),$$

$$\frac{d}{dr}(r^2 L h) = r^2 \int_0^r G_a(r') \, dr'.$$

We take again the integral of the obtained equality and in the right part we integrate by parts:

$$r^2 L h = \int_0^r r''^2 \underbrace{\int_0^{r''} G_a(r') \, dr'}_{u} \, dr'' = \int_0^{r''} G_a(r') \, dr' \cdot \frac{r''^3}{3} \bigg|_0^r - \int_0^r \frac{r''^3}{3} G_a(r'')dr''$$

$$= \frac{r^3}{3} \int_0^r G_a(r') \, dr' - \int_0^r \frac{r''^3}{3} G_a(r'')dr''.$$

Then the solution for $Lh$ takes the form,

$$Lh = \frac{r}{3} \int_0^r G_a(r') \, dr' - \frac{1}{3} \int_0^r \frac{r'^3}{r^2} G_a(r')dr' = \frac{1}{3} \int_0^r \left( r - \frac{r'^3}{r^2} \right) G_a(r')dr' \equiv G_{1a}(r),$$

where the resulting expression was denoted as $G_{1a}(r)$. The solution of the resulting equation with respect to $h$ has the same form as for $Lh$. Only $G_{1a}(r)$ stands in the integrand instead of $G_a(r)$. The resulting construction allows integration by parts:

$$h(r) = \frac{1}{3}\int_0^r \left( r - \frac{r''^3}{r^2} \right) G_{1a}(r')dr' = \frac{1}{9}\int_0^r \underbrace{\int_0^{r'} \left( r' - \frac{r''^3}{r'^2} \right) G_a(r'')dr''}_{G_{1a}(r')=u}\, d\left( rr' - \frac{r'^4}{4r^2} \right) =$$

$$= \frac{1}{9}\left[ \int_0^{r'} \left( r' - \frac{r''^3}{r'^2} \right) G_a(r'')dr'' \cdot \left( rr' - \frac{r'^4}{4r^2} \right)\Bigg|_{r'=0}^{r'=r} - \right.$$

$$\left. - \int_0^r \left( rr' - \frac{r'^4}{4r^2} \right)\underbrace{\int_0^{r'} \left( 1 + \frac{2r''^3}{r'^3} \right) G_a(r'')dr''}_{u}\, dr' \right] =$$

$$= \frac{r^2}{12}\int_0^r \left( r - \frac{r''^3}{r^2} \right) G_a(r'')dr'' - \frac{1}{9}\left[ \int_0^{r'} \left( 1 + \frac{2r''^3}{r'^3} \right) G_a(r'')dr'' \cdot \left( r\frac{r'^2}{2} - \frac{r'^5}{20r^2} \right)\Bigg|_{r'=0}^{r'=r} - \right.$$

$$\left. - \int_0^r \left( r\frac{r'^2}{2} - \frac{r'^5}{20r^2} \right)\left( 3G_a(r') + \int_0^{r'} \left( -\frac{6r''^3}{r'^4} \right) G_a(r'')dr'' \right)dr' \right] =$$

$$= \int_0^r \left( \frac{r^3}{12} - \frac{r''^3}{12} - \frac{r^3}{20}\left( 1 + \frac{2r''^3}{r^3} \right) + \left( \frac{rr''^2}{6} - \frac{r''^5}{60r^2} \right) \right) G_a(r'')dr'' -$$

$$- \frac{1}{3}\int_0^r \left( \frac{r}{r'^2} - \frac{r'}{10r^2} \right)\underbrace{\int_0^{r'} r''^3 G_a(r'')dr''}_{u}\, dr' =$$

$$= \int_0^r \left( \frac{r^3}{30} - \frac{11r''^3}{60} + \frac{rr''^2}{6} - \frac{r''^5}{60r^2} \right) G_a(r'')dr'' -$$

$$- \frac{1}{3}\left[ \int_0^{r'} r''^3 G_a(r'')dr'' \cdot \left( -\frac{r}{r'} - \frac{r'^2}{20r^2} \right)\Bigg|_{r'=0}^{r'=r} + \int_0^r \left( \frac{r}{r'} + \frac{r'^2}{20r^2} \right) r'^3 G_a(r')dr' \right] =$$

$$= \int_0^r \left( \frac{r^3}{30} - \frac{11r''^3}{60} + \frac{rr''^2}{6} - \frac{r''^5}{60r^2} \right) G_a(r'')dr'' + \frac{7}{20}\int_0^r r''^3 G_a(r'')dr''$$

$$- \frac{1}{3}\int_0^r \left( rr'^2 + \frac{r'^5}{20r^2} \right) G_a(r')dr'.$$

Finally, when $r < \gamma$ we get

$$h(r) = \int_0^r \left( \frac{r^3}{30} + \frac{r'^3}{6} - \frac{rr'^2}{6} - \frac{r'^5}{30r^2} \right) G_a(r')dr'. \qquad (39)$$

Carrying out similar analysis and calculations in the region $\gamma < r < 1$, we get,

$$h(r) = \int_\gamma^r \left[ -\frac{r}{3s^2} + \frac{r'^3}{3r^2s^2} + \left( \frac{r'}{rs^3} - \frac{1}{r^2s^5} \right) \text{sh}(s(r-r')) + \right.$$
$$\left. + \left( \frac{1}{s^4 r} - \frac{r'}{r^2 s^4} \right) \text{ch}(s(r-r')) \right] G_{ab}(r')dr'. \qquad (40)$$

And in the region $r > 1$, we have,

$$h(r) = \int_r^\infty \left( -\frac{r^3}{30} + \frac{rr'^2}{6} - \frac{r'^3}{6} + \frac{r'^5}{30r^2} \right) G_b(r')dr'. \qquad (41)$$

Thus, after adding the general solutions of the corresponding homogeneous equations, which were obtained in [23], to the expressions for velocities and pressures in various regions, we obtain:

for $r < \gamma$

$$u_{ra} = \left\{ -\int_0^r \left( \frac{r^2}{15} + \frac{r'^3}{3r} - \frac{r'^2}{3} - \frac{r'^5}{15r^3} \right) G_a(r')dr' + a_1 + a_2 r^2 \right\} \cos\theta,$$

$$u_{\theta a} = \left\{ \int_0^r \left( \frac{2r^2}{15} + \frac{r'^2}{9} - \frac{r'^3}{6r} - \frac{7r'^5}{90r^3} \right) G_a(r')dr' - a_1 - 2a_2 r^2 \right\} \sin\theta, \qquad (42)$$

$$p_a = \left( 10a_2 r + K_1 r - \frac{C}{\kappa} \frac{\partial \psi_{0a}}{\partial r} \right) \cos\theta + \kappa \int_0^r \psi_{0a} \frac{\partial \psi_{0a}}{\partial r'} dr' + p_{0a};$$

for $\gamma < r < 1$

$$u_{rab} = \left\{ -2\int_\gamma^r \left[ -\frac{1}{3s^2} + \frac{r'^3}{3r^3s^2} + \left( \frac{r'}{r^2s^3} - \frac{1}{r^3s^5} \right) \text{sh}(s(r-r')) + \right. \right.$$
$$\left. + \left( \frac{1}{s^4r^2} - \frac{r'}{r^3s^4} \right) \text{ch}(s(r-r')) \right] G_{ab}(r')dr' +$$
$$\left. + c_1 \left( \frac{\text{ch}(sr)}{s^2r^2} - \frac{\text{sh}(sr)}{s^3r^3} \right) + c_2 \left( \frac{\text{sh}(sr)}{s^2r^2} - \frac{\text{ch}(sr)}{s^3r^3} \right) + \frac{c_3}{r^3} + c_4 \right\} \cos\theta,$$

$$u_{\theta ab} = \left\{ \int\limits_{\gamma}^{r} \left[ -\frac{2}{3s^2} - \frac{r'^3}{3r^3 s^2} + \left( \frac{1}{r^3 s^5} + \frac{1}{rs^3} - \frac{r'}{r^2 s^3} \right) \mathrm{sh}(s(r-r')) + \right. \right.$$

$$\left. + \left( \frac{r'}{r^3 s^4} + \frac{r'}{rs^2} - \frac{1}{s^4 r^2} \right) \mathrm{ch}(s(r-r')) \right] G_{ab}(r') dr' +$$

$$(43)$$

$$+ c_1 \left( \frac{\mathrm{ch}(sr)}{2s^2 r^2} - \frac{\mathrm{sh}(sr)}{2s^3 r^3} - \frac{\mathrm{sh}(sr)}{2sr} \right) + c_2 \left( \frac{\mathrm{sh}(sr)}{2s^2 r^2} - \frac{\mathrm{ch}(sr)}{2s^3 r^3} - \frac{\mathrm{ch}(sr)}{2sr} \right) +$$

$$\left. + \frac{c_3}{2r^3} - c_4 \right\} \sin\theta,$$

$$p_{ab} = \left( \frac{c_3 s^2}{2r^2} - c_4 s^2 r + K_2 r + \frac{K_3}{r^2} - \frac{B}{\kappa} \frac{\partial \psi_{0ab}}{\partial r} + 2\kappa D \, r \int\limits_{\gamma}^{r} \frac{\psi_{0ab}(r')}{r'^4} dr' \right) \cos\theta +$$

$$+ \kappa \int_{\gamma}^{r} \psi_{0ab} \frac{\partial \psi_{0ab}}{\partial r'} dr' + p_{0ab};$$

and for $r > 1$

$$u_{rb} = \left\{ -\int\limits_{r}^{\infty} \left( -\frac{r^2}{15} + \frac{r'^2}{3} - \frac{r'^3}{3r} + \frac{r'^5}{15r^3} \right) G_b(r') dr' + \frac{b_1}{r^3} + \frac{b_2}{r} - b_3 \right\} \cos\theta,$$

$$u_{\theta b} = \left\{ \int\limits_{r}^{\infty} \left( \frac{r'^2}{3} - \frac{r'^3}{6r} - \frac{r'^5}{30r^3} - \frac{2r^2}{15} \right) G_b(r') dr' + \frac{b_1}{2r^3} - \frac{b_2}{2r} + b_3 \right\} \sin\theta, \qquad (44)$$

$$p_b = \left( \frac{b_2}{r^2} + \frac{K_4}{r^2} - \frac{1}{\kappa} \frac{\partial \psi_{0b}}{\partial r} - 2A\kappa r \int\limits_{r}^{\infty} \frac{\psi_{0b}(r')}{r'^4} dr' \right) \cos\theta -$$

$$- \kappa \int_{r}^{\infty} \psi_{0b} \frac{\partial \psi_{0b}}{\partial r'} dr' + p_{0b}.$$

The coefficients $a_i$, $b_i$, $c_i$ are incorporated in the general solutions of the homogeneous equations corresponding to (26), (27), and will be determined in the process of solving the boundary value problem. Similar terms with coefficients $K_i$ are included in the partial solutions of inhomogeneous equations. We find them by substituting the obtained solutions (42)-(44), as well as expressions for the distribution of the electrostatic potential (16)-(21), in equations (26), (27). We get, respectively,

$$K_1 = 6\kappa\rho\sigma \frac{\exp(-\kappa)(1+\kappa) - \exp(-\kappa\gamma)(1+\gamma\kappa)}{2 - 2\gamma^3(\sigma-1)^2 + 5\sigma - 2\sigma^2},$$

$$K_2 = \rho \frac{\{-3\sigma - \gamma\kappa(1+5\sigma) + 3\exp(-2\kappa\gamma)(1+\gamma\kappa)\sigma + 3\left(\exp(-\kappa(1-\gamma)) - \exp(-\kappa(1+\gamma))\right)(1+\kappa)\sigma\}}{\gamma(2 - 2\gamma^3(\sigma-1)^2 + 5\sigma - 2\sigma^2)},$$

$$K_3 = \rho \frac{\{-3 - 6\sigma + 3\gamma^2\kappa^2(1+\sigma) + \gamma^3\kappa^3(2+\sigma) + 3\exp(-\kappa(1-\gamma))(1+\kappa)(1+2\sigma+\gamma^2\kappa^2\sigma - \gamma\kappa(1+2\sigma)) + 3\left(\exp(-2\kappa\gamma)(1+\gamma\kappa) - \exp(-\kappa(1+\gamma))(1+\kappa)\right)\cdot(1+2\sigma+\gamma^2\kappa^2\sigma+\gamma\kappa(1+2\sigma))\}}{2\kappa^2(2 - 2\gamma^3(\sigma-1)^2 + 5\sigma - 2\sigma^2)},$$

$$K_4 = 0. \tag{45}$$

The constants $a_i$, $b_i$, $c_i$ in solutions (42)-(44) are found from the boundary conditions (6), (7), (10) and the conditions for the vanishing of the sum of forces (11) acting on the capsule. As it is easy to establish, it follows from condition (6) that the required dimensionless electrophoretic velocity is equal to the constant $b_3$. There are extra boundary conditions since the condition for normal stresses splits into two: a spherically symmetric part and a polar part proportional to the cosine. Another condition is the pressure value at a substantial distance from the capsule. Formally, to determine the electrophoretic mobility, the spherically symmetric part can be ignored, so that there are nine inhomogeneous linear algebraic equations with constant coefficients for nine unknowns $c_1$, $c_2$, $c_3$, $c_4$, $a_1$, $a_2$, $b_1$, $b_2$ and $b_3$. The coefficients of the mentioned algebraic system are either constants, or hyperbolic functions, or integrals that are not taken in elementary functions, including improper ones. Due to their cumbersome nature, these expressions are not given here.

## 5. Results of calculations and discussion

The solution of the linear algebraic system was performed using standard procedures of the Mathematica®-12 application package of Wolfram Research. It should be noted that significant computational difficulties arose in the calculation process. Even though the solution was obtained completely analytically, and the integrals appearing in it are calculated symbolically, the expressions for the components of the flow velocity are multi-page formulae. They include many terms of different orders and different signs. This effect increased with the growth of the parameter $\kappa$ since this parameter repeatedly appears in the solution as an argument of the exponential function. In particular, the solution includes expressions of the type $x$ - $y$, where both numbers have the order of $10^{100}$ or more. The standard double machine precision is approximately 16 digits. If, for example, the true value of the $x$ - $y$ difference is of the order of 1, then when calculating the difference of approximately equal $x$ and $y$ of the order of $10^{100}$, rounded to 16 digits, a pseudo-random noise is obtained. For this reason, part of the calculations that must be performed using ready-made symbolic formulae, is carried out at the minimum allowable value of $\kappa = 10$. In other parametric studies, the solution was found by numerical integration with the highest possible machine accuracy. Fortunately, Mathematica allows to set the increased accuracy of calculations and perform computations with a given number of mantissa digits in each term of the expression. This allowed us to conduct some computational experiments for the values of the parameter $\kappa$ up to 600, which are sufficient to make sure that the solution reaches its limit value at $\kappa \longrightarrow \infty$. Note that the mentioned limit was found analytically.

All calculations were carried out for the case of a constant dimensionless complex $\rho = \rho_{f0} / \rho_1$, which is essentially the ratio of the charge density of the porous layer to the value of the external electric field strength. Analyzing the expressions for pressure and electric potentials, it is easy to see that the values $\rho_{f0}$ and $\rho_1$ are involved in them as the ratio $\rho$. Therefore, qualitatively, the pressure distribution patterns in the system, the electric field strength, and the streamlines corresponding to a large charge and a large external field will look the same as for a small capsule charge and a small external field, if their ratio has the same value $\rho$. The difference, of course, will affect the value of the electrophoretic velocity.

Figure 3 shows the characteristic pressure distributions in the system at different values of parameter $\rho_1$. Moreover, the pressure values are reduced to dimensional units (bar) by multiplying by the value $10^2 \rho_1^2 \kappa^{-3}$. The equilibrium pressure distribution $p_0$ (Fig. 3a,d), which is established in the system in the absence of an external field, and which compensates for the potential gradient $\psi_0$, depends quadratically on $\rho$. Therefore, in dimensional units, its maximum is determined by the value of $\rho_{f0}^2$. The additive shift, which distinguishes Fig. 3a and Fig. 3d, is a consequence of the reduction of the arbitrary constant pressure $P_{a0} = 1$ used in the calculations to the dimensional form. The additional pressure $p_1$ (Fig. 3b,e), which arises when an external electric field is applied, depends linearly on $\rho$. For this reason, the resulting total pressure distribution (Fig. 3c,f) significantly depends on which field dominates – the capsule's own field or the external one. For small values of $\rho \sim 0.1$, the total pressure is determined mainly by the potential of the external field (Fig. 3d,e,f), for large values of $\rho \sim 10^3$ – by the equilibrium distribution of the potential $\psi_0$ (Fig. 3a,b,c). For the intermediate value $\rho \sim 1$, these two potentials have the same order of magnitude and give comparable contributions to the pressure distribution (Fig. 4). Figure 4 is drawn for $\rho_{f0} = 1$; $\rho_1 = 1$ and would qualitatively look the same for any other relevant values of $\rho_{f0}$ and $\rho_1$. The change would only affect the scale of the pressure. It should be noted that the characteristic values of $\rho_{f0}$ and $\rho_1$ are 1 and 0.001, respectively. So, Fig. 3a,b,c drawn at these values, represent a characteristic picture of the pressure distribution in the electrolyte-capsule system.

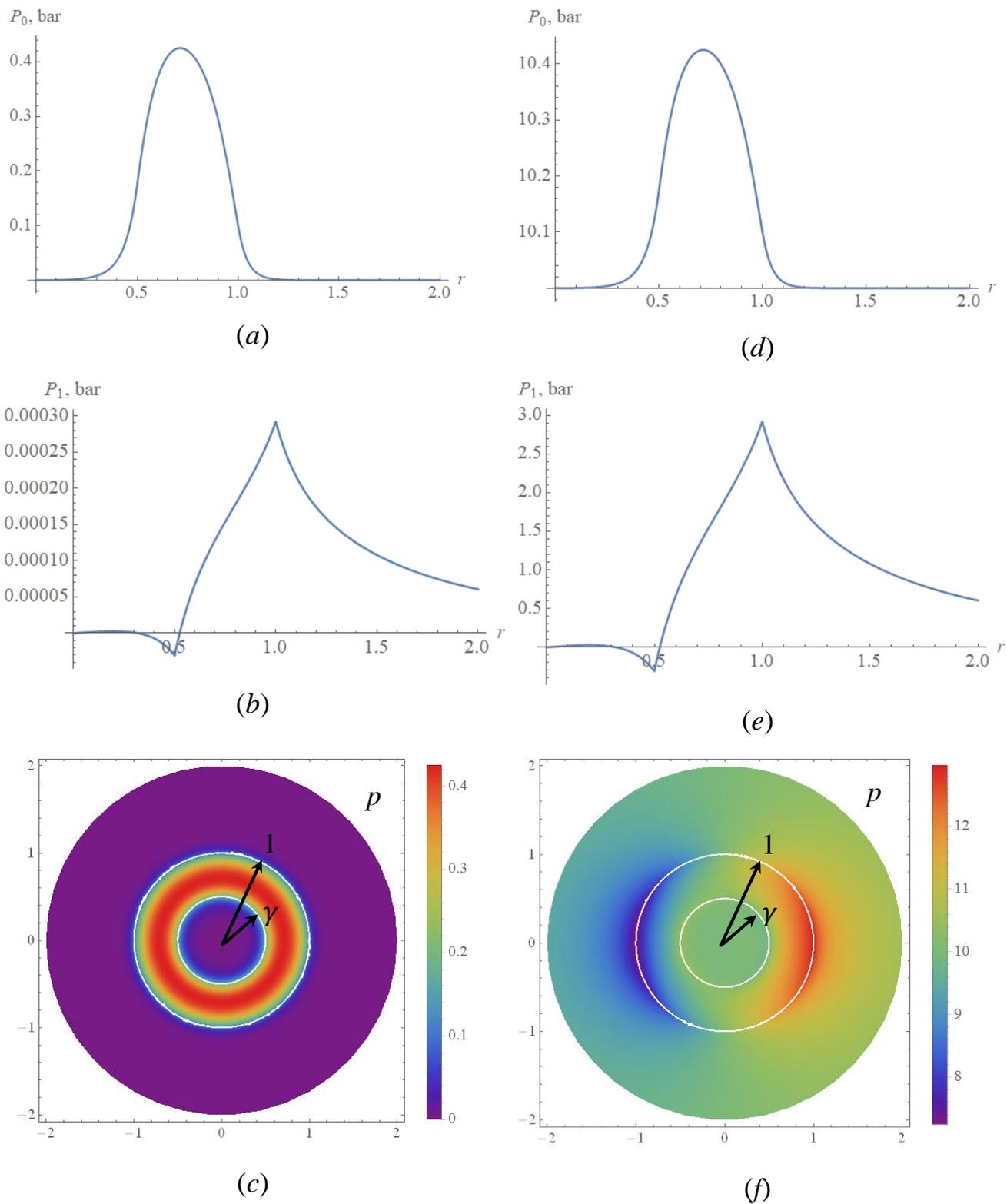

Fig. 3. Distribution of pressure $p_0(r)$, $p_1(r,0)$ and $p(r,\theta)$ for $\gamma = 0.5$; $\kappa = 10$;

$\sigma = 1$; $s = 3$: $(a,b,c) - \rho_{f0} = 1$; $\rho_1 = 0.001$; $(d,e,f) - \rho_{f0} = 1$; $\rho_1 = 10$

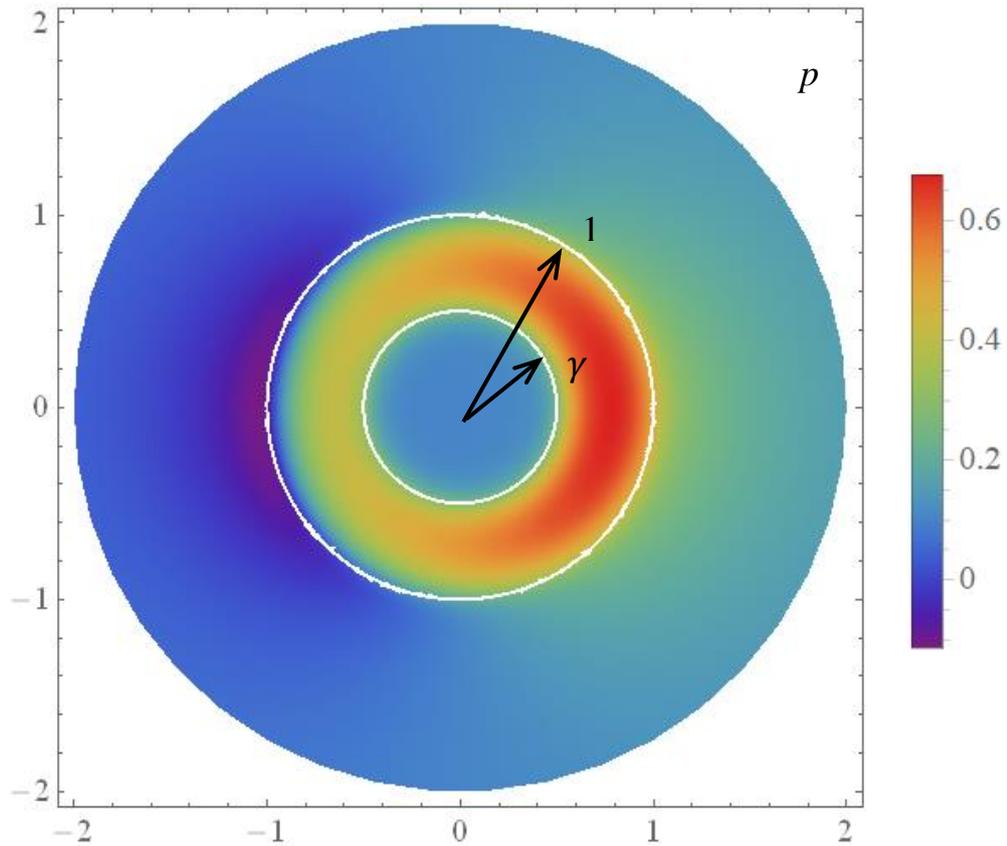

Fig. 4. Pressure $p(r,\theta)$ distribution in the electrolyte-capsule system at $\rho_{f0} = 1$;

$\rho_1 = 1$; $\gamma = 0.5$; $\kappa = 10$; $\sigma = 1$; $s = 3$

Note that the flow pattern is affected only by the pressure distribution $p_1(r,\theta)$. The characteristic distribution of the streamlines is shown in Figure 5a for the following parameter values: $\rho_{f0} = 1$; $\rho_1 = 1$; $\gamma = 0.5$; $\kappa = 10$; $\sigma = 1$; $s = 3$. Of course, the total intensity of the electric field, external and induced by the distribution of charges in the electrolyte and the capsule, significantly depends on the parameter $\rho$: at large value of $\rho \sim 10^3$, a strong perturbation of the field in the vicinity of the capsule is observed (Fig. 5b); at small value of $\rho \sim 0.001$, a uniform external field plays a decisive role (Fig. 6b). The intermediate value of $\rho \sim 1$ gives a similar picture (Fig. 6a), namely, at a sufficient distance from the capsule, the field remains uniform, and in the vicinity of the capsule, the lines of a force field are significantly deformed, and the measure and nature of their deformation are determined by the ratio between $\psi_0$ and $\psi_1$.

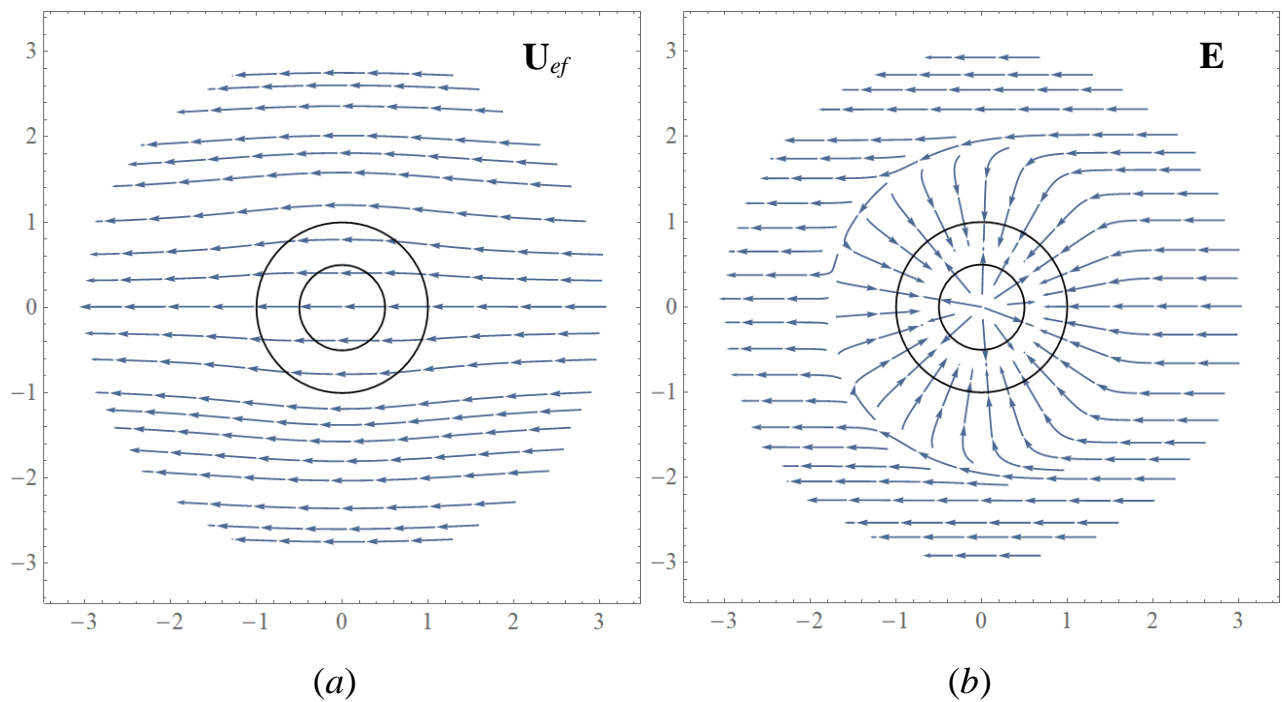

(*a*)                                                 (*b*)

Fig. 5. Flow pattern in the electrolyte-capsule system at $\rho_{I0} = 1$; $\rho_1 = 0.001$; $\gamma = 0.5$;

$\kappa = 10$; $\sigma = 1$; $s = 3$: (*a*) – streamlines; (*b*) – electric field strength lines

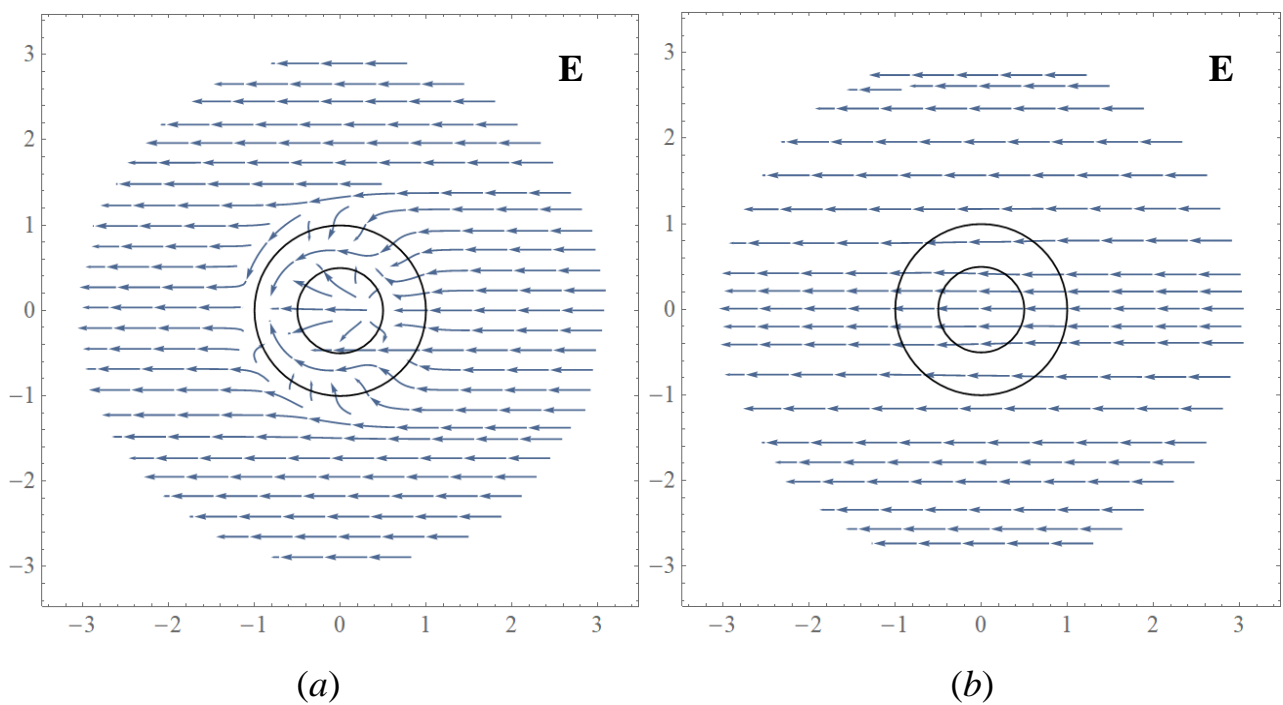

(*a*)                                                 (*b*)

Fig. 6. Electric field strength lines at different values of the capsule charge density

(*a*) – $\rho = 1$; (*b*) – $\rho = 0.001$; $\gamma = 0.5$; $\kappa = 10$; $\sigma = 1$; $s = 3$

Of course, the flow pattern and the electrophoretic velocity, which is the main research subject of this work, depend not only on the parameter $\rho$. Both the thickness of the capsule wall $1 - \gamma$, the capsule reverse permeability $s$, and the conductivity ratio $\sigma$ significantly affect the rate of electrophoresis and may have the opposite effects. The reverse length of the Debye radius $\kappa$ is responsible for the limits of applicability of the model. The larger it is, the more accurately the assumptions made are fulfilled.

The characteristic values of the electrophoretic velocity $U$ are centimeters per second at external fields $E$ of the order of 1 V/cm. So, for the convenience of analyzing the results of calculations, the electrophoretic velocity $U$ will be represented in dimensional units using the relation $U = U_{ef} \cdot \rho_1^2 \cdot \kappa^{-3} \cdot 10^6 \,\mathrm{cm/s}$ , where $U_{ef}$ is the dimensionless value found after solving the problem.

## 5.1. Dependence of the electrophoretic velocity on the reverse permeability s

The parameter $s$ appears in the Brinkman equation and is the ratio of the characteristic size of the problem to the square root of the permeability of the porous layer of the capsule. The characteristic dependences of the electrophoretic velocity on $s$ are shown in Fig. 7 for the following parameter values $\rho_{f0} = 1$; $\rho_1 = 1$; $\sigma = 1$; $\gamma = 0.5$ (Fig. 7a), $\gamma = 0.99$ (Fig. 7b) and $\gamma = 0.2$ (Fig. 7b) and different sets of values $\kappa$.

All the dependencies shown in Figure 7 are smooth for any set of parameters. This confirms the stability of the calculation procedure and the reliability of the results. The series of points obtained at different values of the reverse Debye radius and other equal conditions have an identical form. Moreover, with an increase in the parameter $\kappa$, i.e., a decrease in the thickness of the EDL, the velocity of electrophoresis decreases as expected because electrokinetic phenomena are determined by the EDL and should be demolished when the double layers disappear. With the set of parameters used for the drawing of Fig. 7a,b, a decrease in the permeability of the porous layer (growth of $s$) leads to a drop in the flow velocity. As can be seen from Fig. 5a, there is a complete penetration of the electrolyte into the capsule. There is no pure flow around the particle, as opposed to the case of a solid impermeable particle. Therefore, parameter *s* plays the role of additional resistance to the flow.

A fundamentally different situation is observed in Figure 7c, which shows the dependence of the electrophoretic velocity on the reverse permeability for a capsule with a very thick wall ($\gamma = 0.2$). It is easy to see that all series of points have a pronounced minimum in the vicinity of $s \sim 6$. This means that the same velocity of electrophoretic motion can be achieved in the system at different values of *s*, while all other parameters remain unchanged. For example, at $\kappa = 200$, the electrophoretic velocities corresponding to $s = 3$ and $s = 10$ coincide, although the flow patterns are different: at $s = 3$, the electrolyte penetrates more intensely into the capsule, and at $s = 10$, it almost completely flows around it. Thus, locally different flow patterns can give the same integral characteristic, which is the velocity of electrophoresis. In the first case, there is a more intense flow in a quite sizeable region (the thickness of the porous wall in this example is 0.8), and in the second – a weak flow inside the porous area, but more effective flow around the capsule with an external flow. Thus, two competing effects can lead to the same integral result.

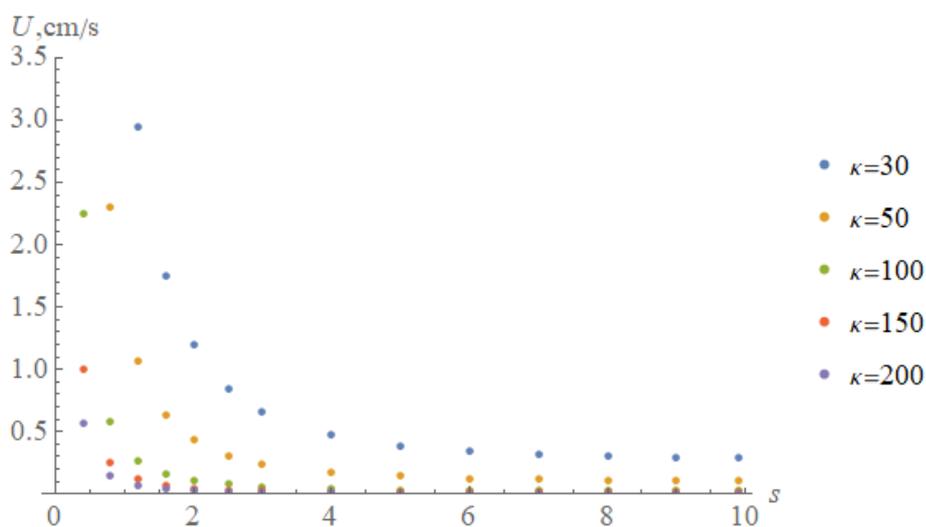

($a$)

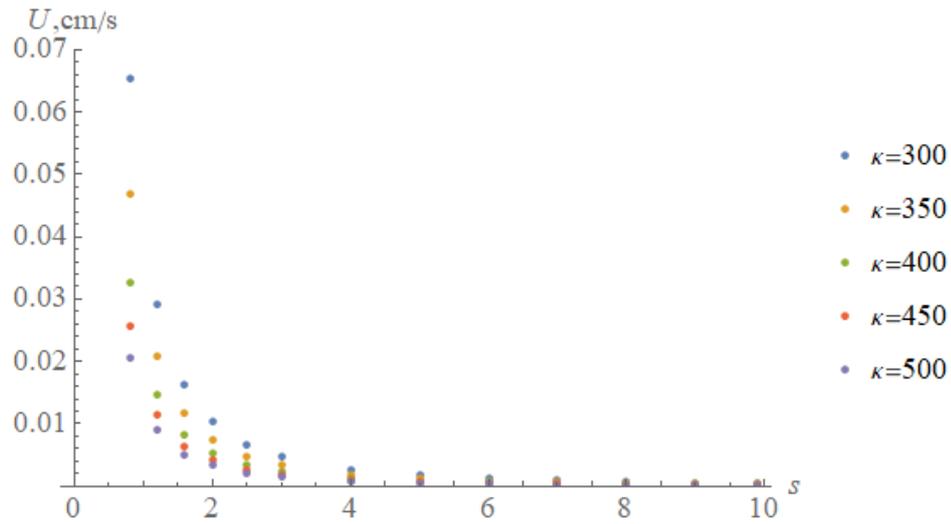

(*b*)

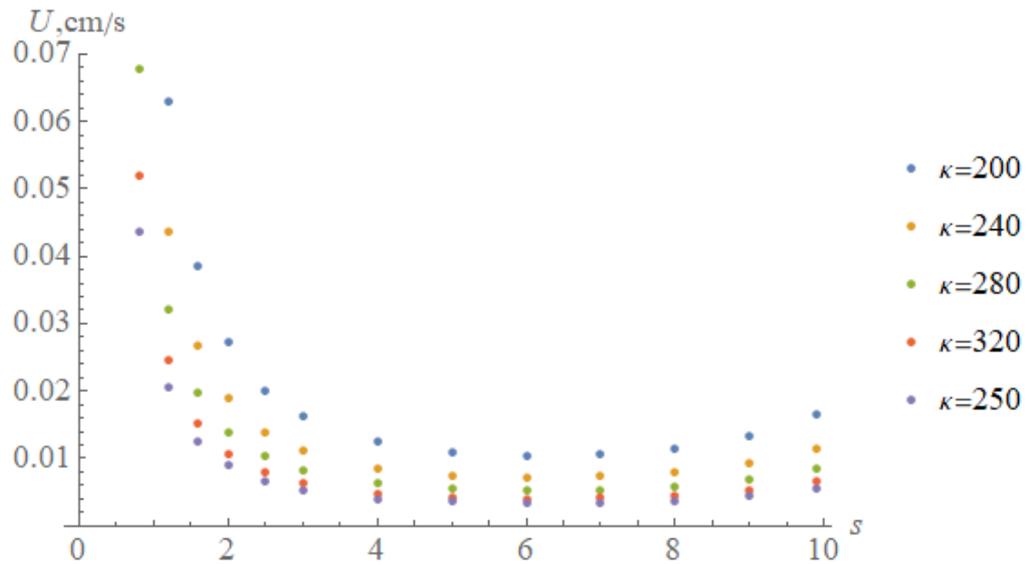

(*c*)

Figure 7. Dependence of the electrophoretic velocity on the reverse permeability of the capsule *s*: (*a*) − $\gamma = 0.5$; (*b*) − $\gamma = 0.99$; (*c*) − $\gamma = 0.2$; $\rho_{f0} = 1$; $\rho_1 = 1$; $\sigma = 1$

The revealed effect is more pronounced the larger the area where the flow depends on *s*, i.e., the thicker the porous wall of the capsule. For thin-walled capsules ($\gamma = 0.99$), with an increase in *s*, a monotonous drop in the electrophoretic velocity is observed up to zero, as in Fig. 7b, since the influence of this parameter affects an exceedingly small region and is reduced to flow inhibition. An intermediate pattern is observed for capsules with an average wall thickness, for example, $\gamma = 0.5$. Figure 7a shows that, starting from a certain *s*, the electrophoretic velocity reaches a non-zero constant and depends on the parameters of the problem, in particular, on $\kappa$. Thus, for capsules with an average wall thickness, the electrophoretic velocity ceases to depend on $s$, starting from a certain value. At the same time, the flow patterns corresponding to different values of $s$ remain different. The revealed effect will be discussed once again when studying the dependence of the electrophoresis rate on the thickness of the capsule wall.

## 5.2. Dependence of the electrophoretic velocity on the reverse Debye radius $\kappa$

The parameter $\kappa$ determines the limits of applicability of the model, since the value $1/\kappa$ characterizes the thickness of the EDL adjacent to each interface of the phases. There are two such interfaces in this problem: $r = \gamma$ and $r = 1$, and the thickness of the EDL should be much less than the thickness of each flow region. Thus, we obtain the following constraints: for the inner part of the capsule $1/\kappa \ll \gamma$, for the porous wall of the capsule $2/\kappa \ll 1-\gamma$, for the outer flow region $1/\kappa \ll 1$. From the last inequality, it follows that the parameter $\kappa$ should be set to at least 10. The second inequality indicates that working with very thin-walled capsules ($\gamma$ close to unity) will only be correct for exceptionally large values of $\kappa$. For example, if $\gamma = 0.99$, we get the limit $\kappa >> 200$. For this reason, the calculations for drawing Fig. 7b were carried out at $\kappa = 300$ or more at the limit of the computing power of a modern computer. From the first inequality, we can conclude that, when working with small $\gamma$, we should use only $\kappa \gg 1/\gamma$, which is also quite problematic due to the mentioned computational difficulties. At $0.2 < \gamma < 0.8$, moderate values of $\kappa$ can be used in calculations, namely, $\kappa = 30$ and higher. In this case, the overlap of the EDL adjacent to the interface boundaries does not occur, and the model remains physically sound.

Figure 8 shows the characteristic dependences of the electrophoretic velocity on $\kappa$ for a capsule with a thick wall $\gamma = 0.5$ (Fig. 8a) and for a capsule with a thin wall $\gamma = 0.99$ (Fig. 8b) with the following parameter values $\rho_{f0} = 1$; $\rho_1 = 1$; $s = 3$; under a set of values $\sigma = \{0.1; 0.5; 1; 5; 10\}$. In accordance with the observation made regarding the acceptable values of $\kappa$ and $\gamma$, the computations for the thick-walled capsule used moderate values of $\kappa$ from 30 to 200 which are more convenient from the point of view of calculations. For a thin-walled capsule, which is usually synthesized and used in experiments, the dependences were drawn in the range of $\kappa$ from 400 to 600. Both graphs in Fig. 8 show a monotonous dependence of the electrophoretic velocity on $\kappa$, the form of this dependence is preserved at different values of the ratio $\sigma$ of the conductivity of the capsule and the electrolyte.

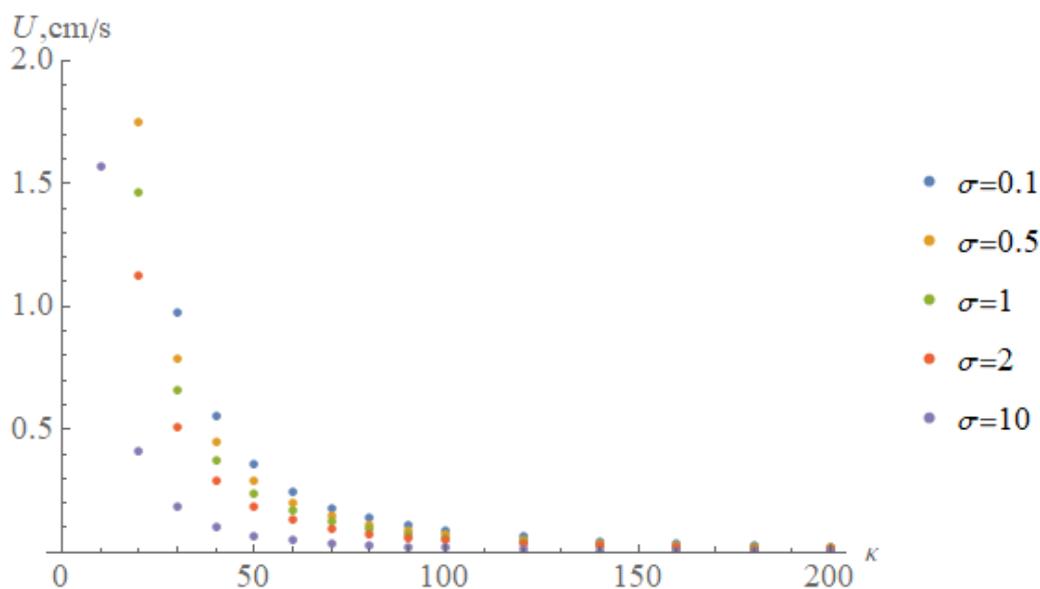

(*a*)

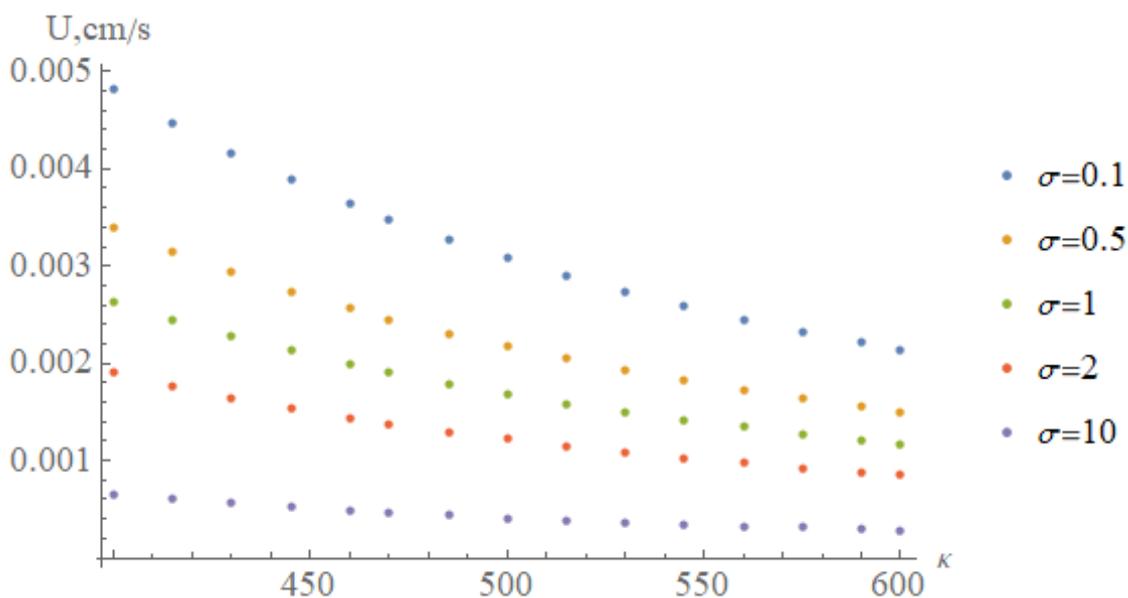

(*b*)

Fig. 8. Dependences of the electrophoretic velocity on the reverse Debye radius $\kappa$:

$(a) - \gamma = 0.5; (b) - \gamma = 0.99; \rho_{f0} = 1; \rho_1 = 1; s = 3$

## 5.3. Dependence of the electrophoretic velocity on the ratio $\sigma$ of electrical conductivities

The parameter $\sigma = \sigma_i / \sigma_e$ was introduced as the ratio of the electrical conductivity of the capsule shell filled with electrolyte to the conductivity of the pure electrolyte. Obviously, its deviation from unity cannot be too large. Therefore, the calculations were based on $\sigma$ values in the interval from 0.1 to 10. The characteristic dependence of the velocity of electrophoresis on $\sigma$ is shown in Fig. 9 for the following values of the parameters $\rho_{f0} = 1$; $\rho_1 = 1$; $s = 3$; $\gamma = 0.5$ (Fig. 9a) and $\gamma = 0.99$ (Fig. 9b); and the corresponding sets of values of $\kappa$.

The drop in the electrophoretic velocity shown in Fig. 9 with the increase in $\sigma$ is quite natural. At higher values of $\sigma$, the electrolyte runs more intensely into the porous region and is sucked by the capsule, as can be seen in Figure 10a, where the velocity field is represented at $\rho_{f0} = 1$; $\rho_1 = 1$; $s = 9$; $\gamma = 0.5$ and $\sigma = 23$. As a result, the flow velocity in the outer region decreases. At even higher values of $\sigma$, vortices can form at some distance from the capsule, that leads to the initiation of return flows. At extremely high values of $\sigma$, the vortices shift towards the capsule and capture the flow in it. However, in practice, such situations can hardly be realized, since they correspond to the conductivity of the capsule wall impregnated with electrolyte which greatly exceeds the conductivity of the pure electrolyte. At $\sigma < 1$, on the contrary, the liquid is forced out of the capsule and flows around it at a higher rate, as can be seen in Figure 10b. The velocity field in Fig. 10b corresponds to the following parameter values: $\rho_{f0} = 1$; $\rho_1 = 1$; $s = 9$; $\gamma = 0.5$ and $\sigma = 0.001$.

One more feature of the dependence of the electrophoretic velocity on the conductivity ratio $\sigma$ is shown in Fig. 9: in contrast to the dependence $U(s)$ (Fig. 7), the dependence of the electrophoretic velocity does not have an asymptotic growth in the vicinity of $\sigma = 0$. On the contrary, when $\sigma \to 0$ the electrophoretic velocity tends to a finite magnitude, which is determined by the values of other parameters. It means that a fully dielectric capsule can also be used for electrophoresis. At the same time, its velocity will be even greater than that of the conducting capsule, all other characteristics being equal.

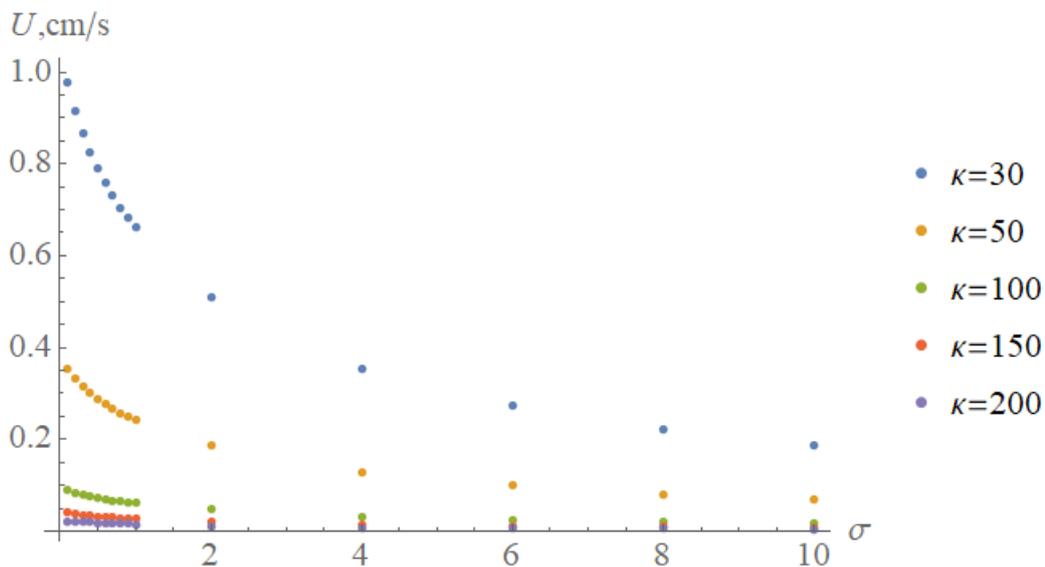

(*a*)

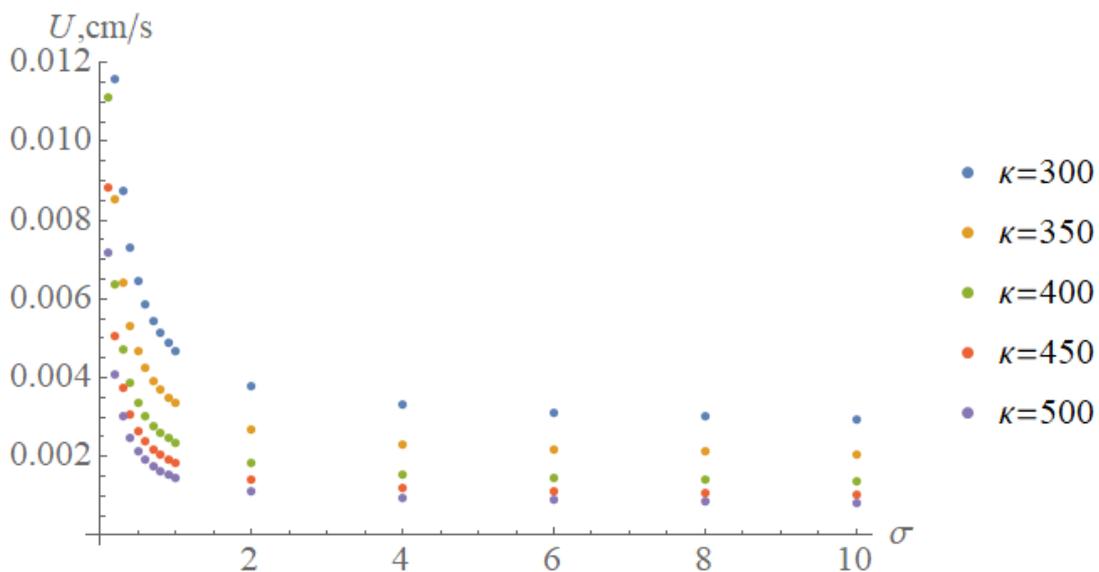

(*b*)

Fig. 9. Dependences of the electrophoretic velocity on the conductivity ratio $\sigma$: (*a*) – $\gamma = 0.5$; (*b*) – $\gamma = 0.99$; $\rho_{f0} = 1$; $\rho_1 = 1$; $s = 3$

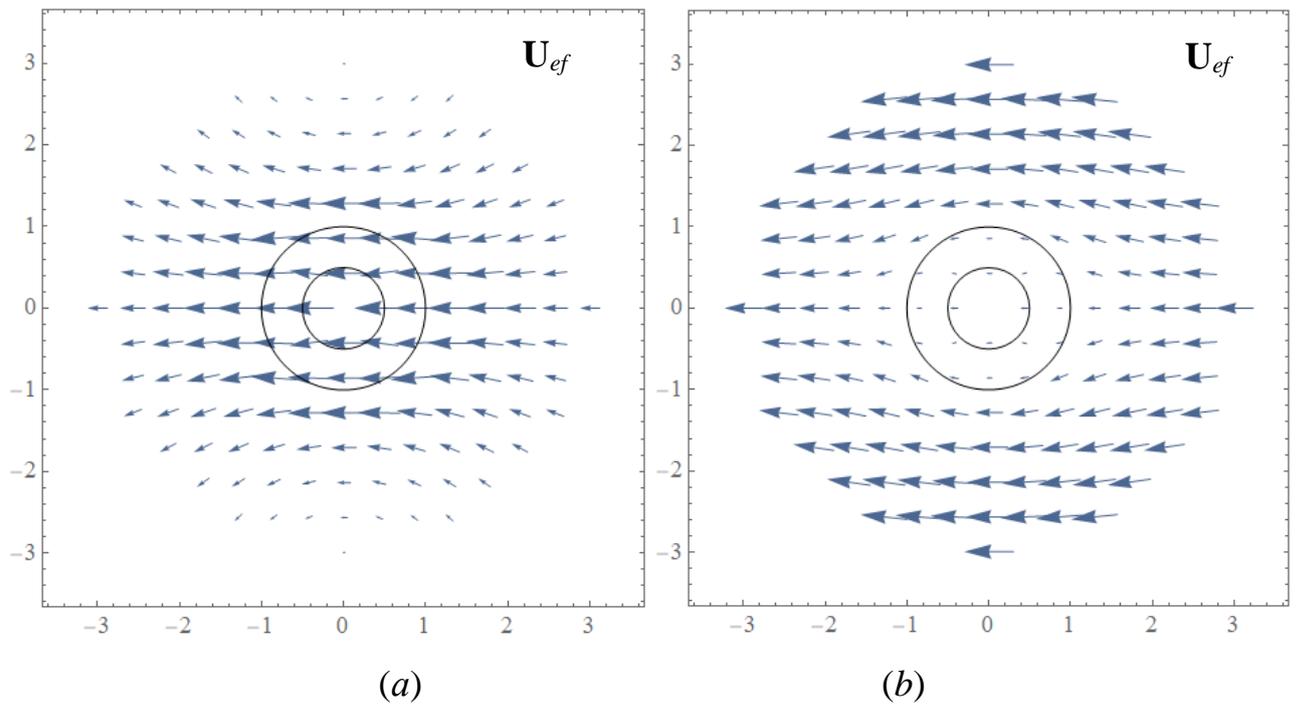

Fig. 10. Velocity field in the electrolyte-capsule system at $(a) - \sigma = 23$, $(b) - \sigma = 0.001$; $\rho_{f0} = 1$; $\rho_1 = 0.001$; $\gamma = 0.5$; $\kappa = 10$; $s = 9$

### 5.4. The dependence of the electrophoretic velocity on the thickness of the capsule wall $1 - \gamma$

Even though the results only for thin-walled capsules are of practical significance, an analysis of the solution for the entire range of values of $\gamma$ will be carried out here for a more complete understanding of the developed model and the specifics of the problem solution. Figure 11 shows the dependence of the electrophoretic velocity on the parameter $\gamma$, which varies in the interval $(0.2; 0.8)$. As mentioned above, for this interval, it is acceptable to use moderate values of $\kappa$, which do not lead to serious computational difficulties. When drawing the dependencies, the following values of the parameters were used: $\rho_{f0} = 1$; $\rho_1 = 0.001$; $\kappa = 50$; $\sigma = 1$ at different values of parameter $s$. Figure 11b shows part of Figure 11a and contains the dependencies $U(\gamma)$ only for $s = 3$, 5, and 10, which are located close to the abscissa axis and are not clearly visible in Figure 11a.

It should be mentioned that the resulting dependencies, although monotonic, have derivatives of different signs. At low values of $s$, the electrophoretic velocity increases as the thickness of the capsule wall decreases. The reason for this effect may be that small values of $s$ create conditions of increased permeability to the liquid. The larger the volume fraction of such a flow region (i.e., the smaller $\gamma$), the more it draws liquids into the capsule and the lower the resulting velocity in the external flow, i.e., the velocity of electrophoresis. For sufficiently large values of $s$ (see Figure 11b), the situation is exactly opposite. The porous region has an increased resistance to the flow, pushing it out. The greater the flow rate of the capsule, the more efficiently it is done, i.e., the larger the flow area involved in this process (the smaller $\gamma$). Obviously, there is the value of $s$ at which the electrophoretic velocity remains unchanged at any magnitude of $\gamma$. Of course, this value of $s$ depends on $\kappa$ and $\sigma$, i.e., on the concentration of the electrolyte and the conductive properties of the polyelectrolyte using which the capsule is synthesized. However, the revealed feature of the $U(\gamma)$ dependence means that under certain conditions, the thickness of the capsule wall does not matter – the velocity of electrophoresis will depend on other parameters.

Another unique feature of the dependencies $U(\gamma)$ shown in Figure 11b is the presence of the intersection point of the curves corresponding to different values of $s$. Undoubtedly, under different $s$ and all other equal conditions, the flow patterns will be different. But, as it turns out, the velocity at infinity can be the same. This conclusion was already obtained when considering the $U(s)$ dependencies in Fig. 7. Since for values of $\gamma$ that not close to unity, there are such values of $\kappa$ and $\sigma$ at which the electrophoretic velocity is constant, $U(s) = \text{const}$, starting from some $s$, the presence of points of intersection of the dependencies $U(\gamma)$ at different $s$ is quite predictable. From a practical point of view, this means that a given velocity of electrophoresis of a polyelectrolyte capsule can be achieved by using materials with different permeability and other matching characteristics for its manufacture.

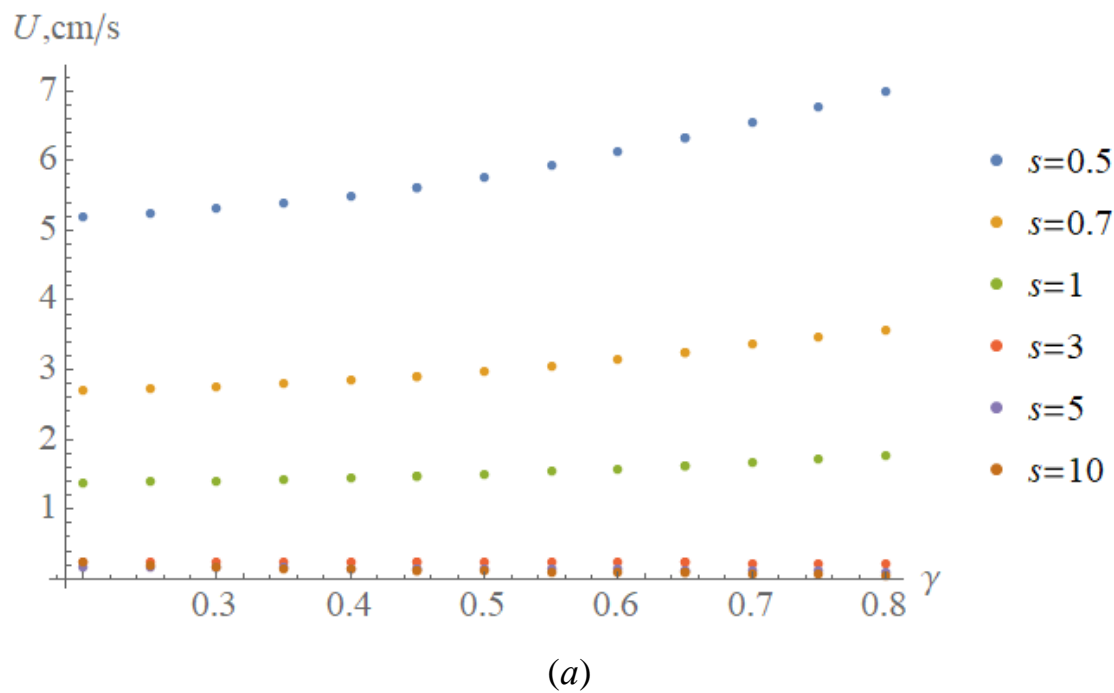

(*a*)

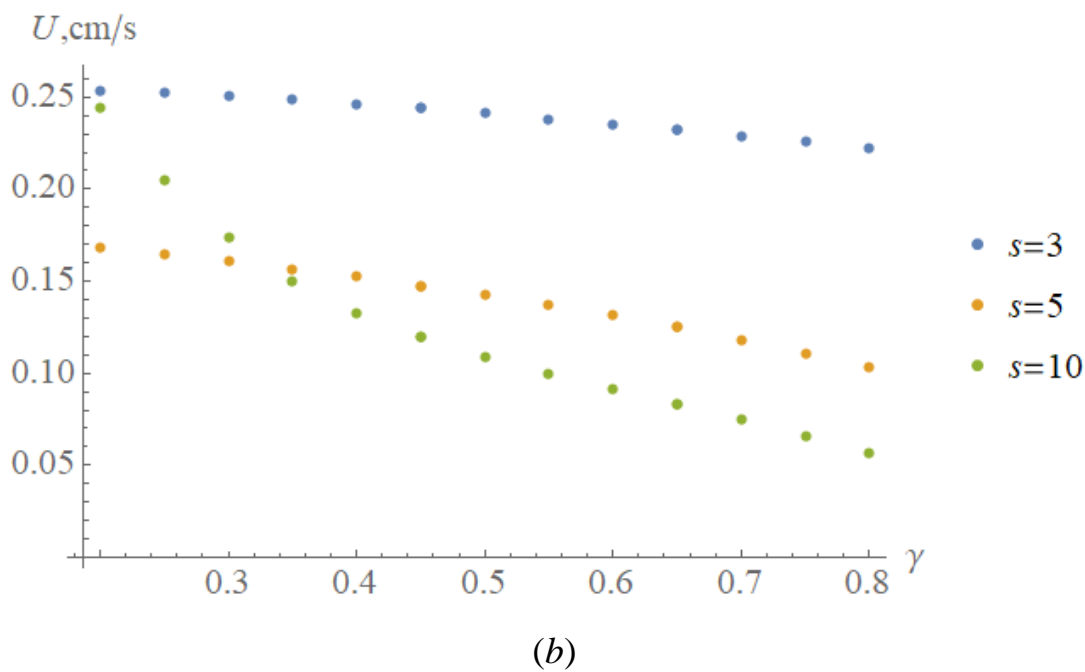

(*b*)

Fig. 11. Dependence of the electrophoretic velocity on the position of the inner boundary of the porous layer $\gamma$; $\rho_{f0} = 1$; $\rho_1 = 1$; $\kappa = 50$; $\sigma = 1$

Finally, we consider the dependencies of $U(\gamma)$ over the entire range of $\gamma$ values allowed within the framework of the model used. The corresponding series of points are shown in Fig. 12 at $\rho_{f0} = 1$; $\rho_1 = 1$; $\kappa = 250$; $\sigma = 1$ and the same values of $s$ as in Fig. 11. The parts of Figure 12 (a) and (b) are similarly organized.

It is easy to see that all the peculiarities revealed in Fig. 11 take place in Fig. 12, with the only difference that they are found at different values of the parameters, and the values of the electrophoretic velocity themselves are lower, since all the dependencies shown are depicted at a significantly higher value of $\kappa$ than in Fig. 11. Using such a large and inconvenient value for calculations allowed us to obtain additional points for small (close to zero) and large (close to unity) values of $\gamma$, which are important for practice. The additional points did not change the nature of the dependence $U(\gamma)$, but could not be obtained at $\kappa = 50$, which was used to draw Fig. 11.

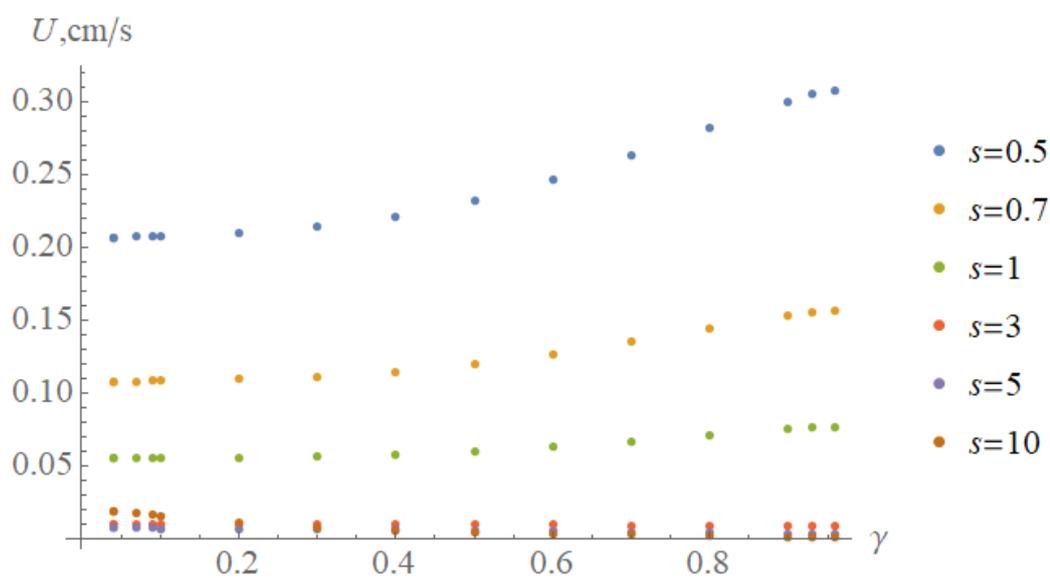

(*a*)

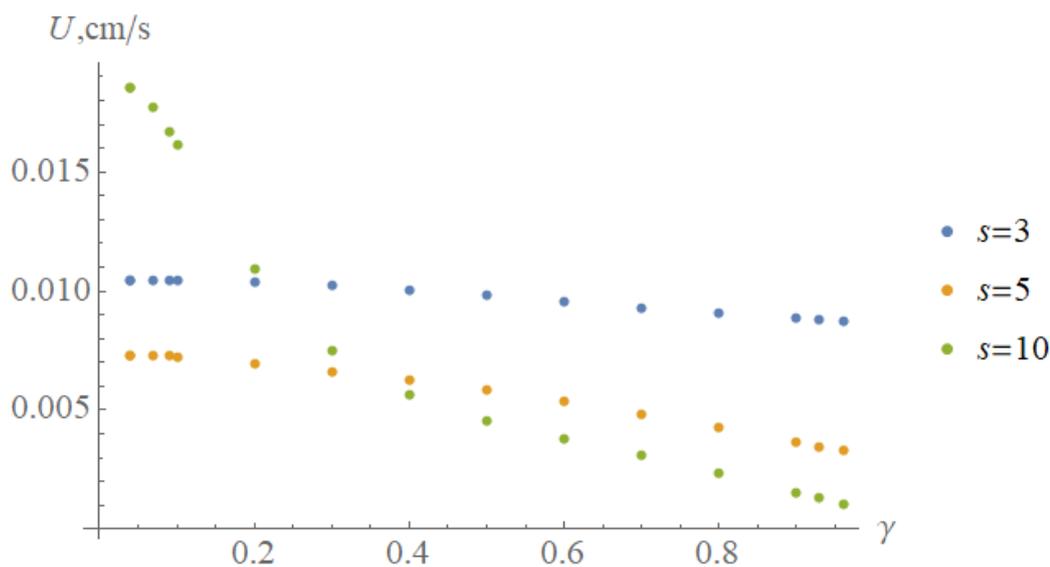

(*b*)

Fig. 12. Dependence of the electrophoretic velocity on the position of the inner boundary of the porous layer $\gamma$; $\rho_{f0} = 1$; $\rho_1 = 1$; $\kappa = 250$; $\sigma = 1$

## 6. Conclusions

In this paper, we considered the problem of electrophoresis of a polyelectrolyte capsule with a uniformly charged, porous conducting shell which moves under the influence of an external electric field in an electrolyte that is identical to the one contained in the capsule's cavity.

An exact analytical solution of the corresponding boundary value problem was obtained for the case of small electrical potentials. The solution was studied numerically for different values of the specific permeability of the capsule, and the thickness of the porous and the electric double layer.

The calculations showed that the velocity of electrophoresis of a porous capsule depends on the parameters of the capsule and the characteristics of the dispersion medium in a rather complex way. The velocity of electrophoresis may either increase or decrease with the thickness of the capsule shell. The pattern of this dependence is determined by the combination of the parameters that characterize the capsule (its conductivity, charge, hydrodynamic permeability), as well as the parameters of the dispersion medium. A minimum on the curve of the electrophoretic velocity vs. the reverse permeability of the porous layer was found. From a practical point of view, this minimum means that a given electrophoretic velocity can be achieved by using materials with various permeability and other matching characteristics when manufacturing a polyelectrolyte capsule. It was shown that electrophoretic mobility decreases together with the conductivity of the porous layer material. This implies that a dielectric capsule can also be used for electrophoresis. At the same time, its velocity will even be greater than that of the conducting capsule, all other parameters being equal. We illustrated these results with graphs, pictures of the pressure distribution in the capsule and in its vicinity, as well as streamlines and electric field lines.

Our study suggests that the fine structure of capsules can be determined by measuring their electrophoretic mobility, as well as their fractionation in an electric field. In addition, based on the obtained results, it is possible to predict the behavior of capsules with known properties under specific conditions, as well as to develop recommendations on the properties of synthesized capsules for their usage in electrophoresis with specified parameters.

## CRediT authorship contribution statement

Anatoly N. Filippov: Conceptualization, Methodology, Supervision, Funding acquisition, Writing - review & editing. Daria Yu. Khanukaeva: Methodology, Formal Analyses, Validation, Software, Data Curation, Writing - original draft. Petr A. Aleksandrov: Methodology, Software, Data curation.

## Declaration of Competing Interest

The authors declare that they have no known competing financial interests or personal relationships that could have appeared to influence the work reported in this paper.

## Acknowledgements

We dedicate this work to the memory of Professor Vyacheslav Roldughin, an outstanding physicochemist and our dear colleague who began working on the problem tackled in this paper about eight years ago but passed away prematurely.

This work was supported by the Russian Science Foundation (project No. 20-19-00670).